\documentclass[a4paper,twocolumn,10pt,accepted=2023-08-01]{quantumarticle}
\pdfoutput=1

\usepackage[utf8]{inputenc}
\usepackage{bbold}
\usepackage{amsmath, amssymb, graphicx, bm}
\usepackage{physics}

\usepackage[labelformat=simple]{subcaption}

\usepackage{xcolor}
\usepackage{hyperref}
\usepackage[capitalise]{cleveref}
\usepackage[numbers,sort&compress]{natbib}

\usepackage{tikz}
\newcommand\Plus[1][1]{%
\begin{tikzpicture}[scale=#1]
\draw[line width=0.3ex, x=1ex, y=1ex] (0.5,0) -- (0.5,1)(0,0.5) -- (1,0.5);
\end{tikzpicture}}

\newcommand{\cutoffs}{\textit{cutoffs}}
\newcommand{\diag}{\textit{diag}}
\newcommand{\offset}{\textit{offset}}
\newcommand{\dvar}{K}
\newcommand{\sumvar}{S}
\newcommand{\C}{\widetilde{C}}

\usepackage{float}
\makeatletter
\let\newfloat\newfloat@ltx
\makeatother
\usepackage{algorithm}
\usepackage[noend]{algpseudocode}
\floatname{algorithm}{Algorithm } % The fix for the float package removes the space between 'Algorithm' and the number in the caption

\makeatletter
\let\OldStatex\Statex
\renewcommand{\Statex}[1][3]{%
  \setlength\@tempdima{\algorithmicindent}%
  \OldStatex\hskip\dimexpr#1\@tempdima\relax}
\makeatother

\algnewcommand{\LeftComment}[1]{\Statex \(\triangleright\) #1}

\algrenewcommand\algorithmiccomment[1]{\hfill // #1}

\newcommand\Algphase[1]{%
\vspace*{-.5\baselineskip}\Statex[0]\hspace*{\dimexpr-\algorithmicindent-2pt\relax}\rule{\textwidth}{0.4pt}%
\vspace*{-.28\baselineskip}%
\Statex[0]\hspace*{-\algorithmicindent}\text{#1}%
\vspace*{-.7\baselineskip}\Statex[0]\hspace*{\dimexpr-\algorithmicindent-2pt\relax}\rule{\textwidth}{0.4pt}%
\vspace*{-.1\baselineskip}%
}
\newcommand{\ts}{\textsuperscript}

\begin{document}

\title{A Quadratic Speedup in the Optimization of Noisy Quantum Optical Circuits}

\author{Robbe De Prins}
\email{robbe.deprins@ugent.be}
\affiliation{Photonics Research Group, INTEC, Ghent University – imec, Sint-Pietersnieuwstraat 41, 9000 Ghent, Belgium}

\author{Yuan Yao}
\email{yuan.yao@telecom-paris.fr
}
\affiliation{T\'el\'ecom Paris and Institut Polytechnique de Paris, LTCI, 20 Place Marguerite Perey, 91120 Palaiseau, France }

\author{Anuj Apte}
\email{apteanuj@uchicago.edu}
\affiliation{Xanadu, Toronto, ON, M5G 2C8, Canada}
\affiliation{Kadanoff Center for Theoretical Physics \& Enrico Fermi Institute, Department of Physics, University of Chicago, Chicago, IL 60637}

\author{Filippo M. Miatto}
\email{filippo@xanadu.ai}
\affiliation{T\'el\'ecom Paris and Institut Polytechnique de Paris, LTCI, 20 Place Marguerite Perey, 91120 Palaiseau, France }
\affiliation{Xanadu, Toronto, ON, M5G 2C8, Canada}

\begin{abstract}
    Linear optical quantum circuits with photon number resolving (PNR) detectors are used for both Gaussian Boson Sampling (GBS) and for the preparation of non-Gaussian states such as Gottesman-Kitaev-Preskill (GKP), cat and NOON states. They are crucial in many schemes of quantum computing and quantum metrology. Classically optimizing circuits with PNR detectors is challenging due to their exponentially large Hilbert space, and quadratically more challenging in the presence of decoherence as state vectors are replaced by density matrices.
    To tackle this problem, we introduce a family of algorithms that calculate detection probabilities, conditional states (as well as their gradients with respect to circuit parametrizations) with a complexity that is comparable to the noiseless case. As a consequence we can simulate and optimize circuits with twice the number of modes as we could before, using the same resources.
    More precisely, for an $M$-mode noisy circuit with detected modes $D$ and undetected modes $U$, the complexity of our algorithm is $O(M^2 \prod_{i\in U} C_i^2 \prod_{i\in D} C_i)$, rather than $O(M^2 \prod_{i \in D\cup U} C_i^2)$, where $C_i$ is the Fock cutoff of mode $i$.
    As a particular case, our approach offers a full quadratic speedup for calculating detection probabilities, as in that case all modes are detected.
    Finally, these algorithms are implemented and ready to use in the open-source photonic optimization library \texttt{MrMustard} \cite{MrMustard_github}.
\end{abstract}

\maketitle

\begin{figure}
\centering
\begin{subfigure}{1\linewidth}
  \includegraphics[width=\linewidth]{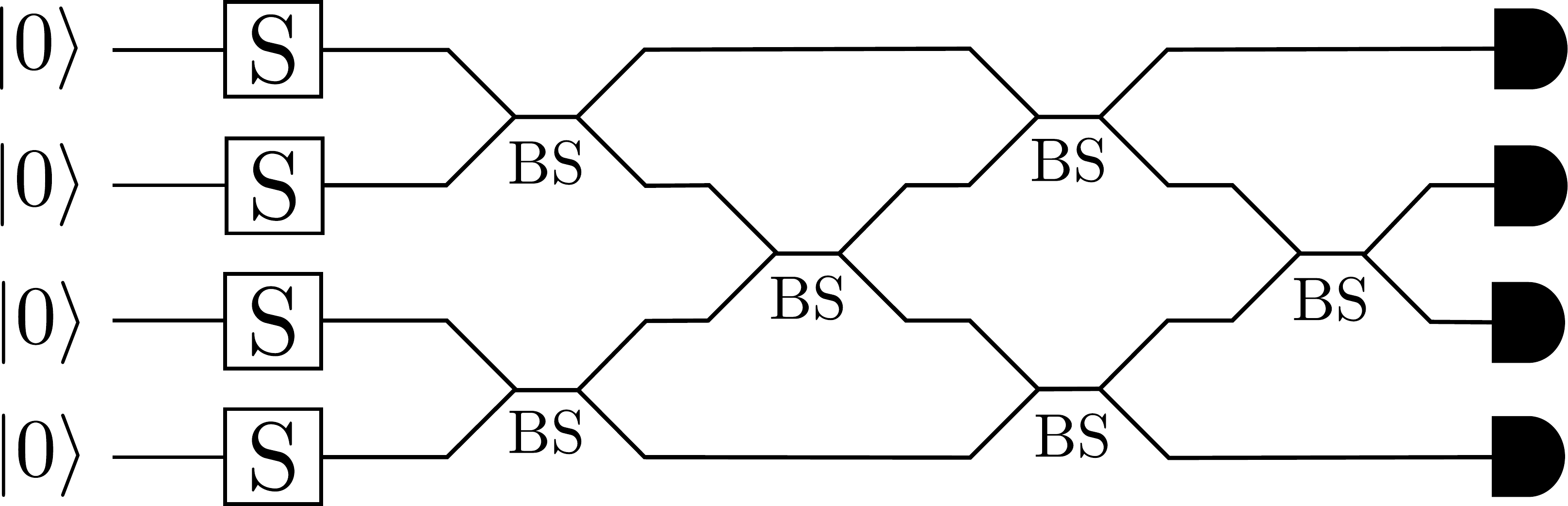}
  \caption{Gaussian Boson Sampling}
  \label{fig:circuit_detectAllModes}
\end{subfigure}
\begin{subfigure}{1\linewidth}
  \includegraphics[width=\linewidth]{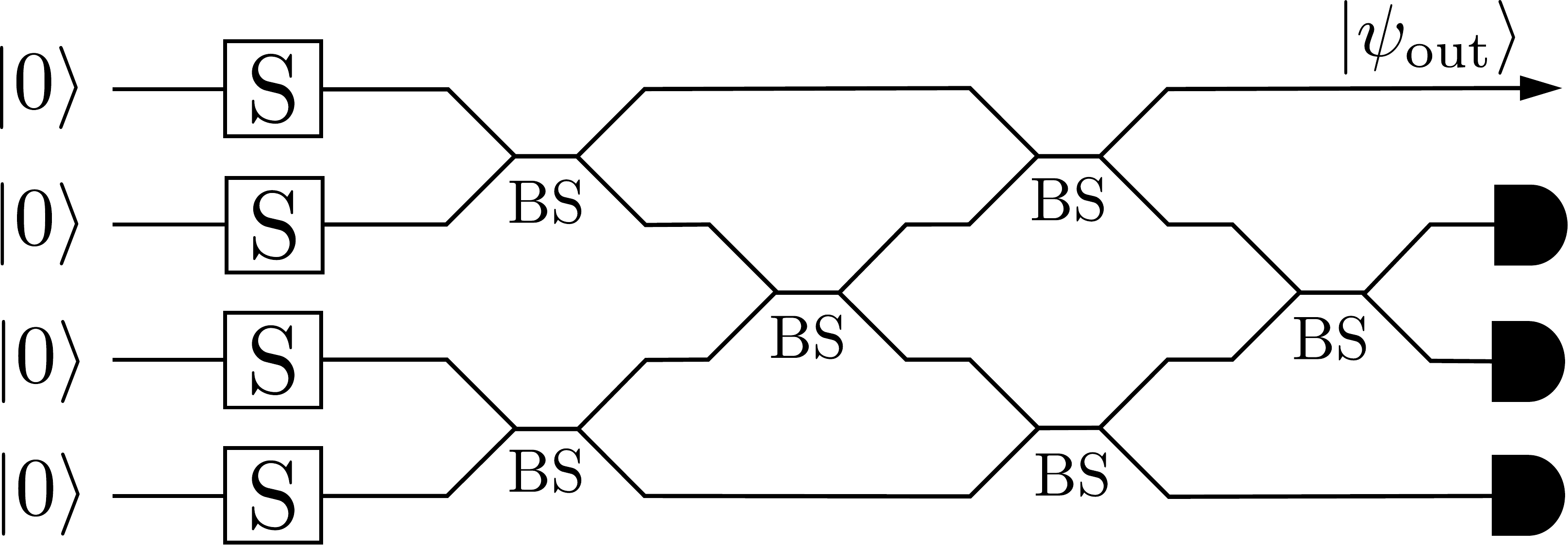}
  \caption{Conditional non-Gaussian state generation 
  \\ (e.g.~GKP, cat, NOON states, etc.)
  }
  \label{fig:circuit_leftovermode}
\end{subfigure}
\caption{Examples of linear optical quantum circuits with PNR detectors. Vacuum states are squeezed and sent through an interferometer. A subset of the modes is measured with PNR detectors.}
\label{fig:example_circuits}
\end{figure}

\section{Introduction}
Linear optical quantum circuits with photon number resolving (PNR) detectors are studied because of two main reasons. First of all, they are used to perform Gaussian Boson Sampling (GBS). In GBS, squeezed states are sent through an interferometer and subsequently detected by PNR detectors. An example of such a circuit is depicted in \cref{fig:circuit_detectAllModes}.

GBS is a leading approach in pursuing quantum advantage \cite{GBS1,GBS2}. Moreover, several quantum algorithms based on GBS have been introduced \cite{bromley2020applications, Arrazola2018a,Arrazola2018b,Banchi2020,Bradler2018,Bradler2021,Jahangiri2020,Schuld2020,Huh2017}, some of which rely on the ability to train the circuit parameters. 

The second (and arguably more useful) application for circuits with PNR detectors is the generation of conditional non-Gaussian states. 
Examples of such states include Gottesman-Kitaev-Preskill (GKP) states, cat states, bosonic-code states, weak cubic phase states, ON states and NOON states \cite{PhysRevA.101.032315, PhysRevA.100.052301, PhysRevA.100.012326, su2019generation, takase2022gaussian, PhysRevLett.128.240503, PhysRevA.103.013710, PhysRevA.100.022341}.
These states are used in a wide range of applications, such as generating bosonic error correction codes, providing resource states for the implementation of non-Gaussian gates and quantum metrology.
We emphasize the particular interest of GKP states \cite{GKP_original} as they are one of the leading candidates for qubits in optical quantum computation \cite{bourassa2021blueprint}.
\cref{fig:circuit_leftovermode} depicts a circuit that can be used to generate non-Gaussian states.
Depending on the PNR detection pattern, a certain state is generated. The probability distribution of all conditional states is governed by the circuit parameters. 
By training these parameters, we can increase the probability of generating certain non-Gaussian states of interest and their quality.

In this work we address these simulation and optimization tasks using the framework that we introduced in our previous work \cite{miatto2020fast,yao2022recursive}. This framework allows one to recursively calculate elements of the matrix representation of Gaussian operators in Fock space. Here, it provides us with the matrix elements that define the detection probabilities or the amplitudes of conditional states. Moreover, we can recursively calculate the gradients of these elements with respect to a circuit parametrization, which allows us to find the parameters that minimize a certain cost function using gradient descent.

In realistic settings, decoherence effects such as photon loss affect the output of quantum circuits. Consequently, we need to be able to include these effects into our simulations if we want them to be faithful and useful. This motivates us to carry out simulations using density matrices. Normally, swapping state vectors for density matrices would make tasks quadratically more demanding in terms of both memory and runtime. We will show that we can almost completely get around this quadratic increase by introducing an algorithm that allows us to apply the recurrence relations fewer times while still including the amplitudes of interest. The resulting algorithm works for circuits with in principle an arbitrary number of PNR detectors. We will show that the complexity of our algorithm is comparable to the complexity of the lossless case, as long as the number of detected modes is a large fraction of the total number of modes. 

The paper is structured as follows. In \cref{sec:recrel and state vectors} we recall our simulation and optimization framework \cite{miatto2020fast,yao2022recursive} and apply it to lossless circuits with PNR detectors (i.e.~using state vectors). 
In \cref{sec:algos density matrices} we extend the framework to density matrices. We do this for GBS circuits (such as \cref{fig:circuit_detectAllModes}) in \cref{sec: diagonal algo} and for conditional state generator circuits (such as \cref{fig:circuit_leftovermode}) in \cref{sec: leftover algo}. In \cref{sec:complexity} we discuss the complexity of our algorithms.
\cref{sec:space_and_time} gives numerical results for the memory requirements and speed. \cref{sec:comparison_bristol} gives a comparison with the state-of-the-art classical GBS simulation method.

Note that the construction of good ansätze for GKP generating circuits, as well as the construction of associated cost functions and target states is a separate research question in itself that we do not address in this manuscript.

\section{Circuit optimization framework revisited}
\label{sec:recrel and state vectors}
\subsection{Representing Gaussian operators in Fock space}

In Reference \cite{miatto2020fast}, it was shown that quantum optical circuits can be simulated by using a recurrence relation that calculates elements of the matrix representation of Gaussian operators (i.e.~pure Gaussian states, mixed Gaussian states, Gaussian unitary transformations or Gaussian channels) in Fock space. 
We will denote such a matrix representation by $\bm{\mathcal{G}}$ and call its elements the `Fock amplitudes' of a Gaussian operator. 

As we are interested in calculating detection probabilities and possible conditional states here, we will consider $\bm{\mathcal{G}}$ to be the matrix representation of the multi-mode Gaussian state before the detectors. In other words, $\bm{\mathcal{G}}$ is either a state vector or density matrix in Fock space.
We represent $\bm{\mathcal{G}}$ as a multidimensional array and refer to its total number of dimensions (i.e.~indices) as $D$. Hence, a general Fock amplitude can be written as $\mathcal{G}_{\bm{k}}$, where $\bm{k}$ is an integer vector of length $D$.
We will refer to $\bm{k}$ as a `Fock index' of $\bm{\mathcal{G}}$. 
If $\bm{\mathcal{G}}$ is a state vector, we use the convention that every element of $\bm{k}$ corresponds to an optical mode. If $\bm{\mathcal{G}}$ is a density matrix, every \textit{pair} of consecutive elements in $\bm{k}$ corresponds to an optical mode. 
For example, $\bm{k} = [m,n,p,q]$ is a general Fock index for a density matrix on 2 modes, where the indices $m,n$ and $p,q$ respectively correspond with the first and second mode.
The expression for the Fock amplitudes using Dirac notation is $\mathcal{G}_{\bm{k}} = \mathcal{G}_{m n p q} = \bra{m,p} \mathcal{G} \ket{n,q}$.
For a general number of modes $M$, it follows that:
\begin{equation}
    D =
    \begin{cases}
        M, &         \text{if $\bm{\mathcal{G}}$ is a state vector},\\
        2M, &         \text{if $\bm{\mathcal{G}}$ is a density matrix.}
    \end{cases}
    \label{eq:cases_D}
\end{equation}

Fock amplitudes can now be calculated using the following recurrence relation:
\begin{equation}
    \mathcal{G}_{\bm{k} + \bm{1}_i} = \frac{1}{\sqrt{k_i+1}}\left(\mathcal{G}_{\bm {k}} b_i + \sum_{l = 1}^D \sqrt{k_l} \: \mathcal{G}_{\bm{k} - \bm{1}_l} A_{il}\right), \label{eq:recrel}
\end{equation}
where $\bm{1}_i$ is a vector of all zeroes except for a single 1 in the i\ts{th} entry.
Note that Fock indices that contain at least one negative value correspond to a zero Fock amplitude, as negative photon numbers are nonphysical. Hence, the sum over $l$ may contain less than $D$ terms.

The matrix $\bm{A}$ and vector $\bm{b}$ in \cref{eq:recrel} are complex-valued parameters (of size $D \times D$ and $D$ respectively) that are easily acquired for a specific circuit as they derive from the parameters of the Gaussian representation. 
If $\bm{\mathcal{G}}$ is a density matrix $\bm{\rho}$ we recall the results derived in Reference \cite{yao2022recursive} that relate $\bm{A}_{\bm{\rho}}$ and $\bm{b}_{\bm{\rho}}$ to its complex (i.e.~in the $a$/$a^\dagger$ basis) covariance matrix $\boldsymbol{\sigma}$ and displacement vector $\boldsymbol{\mu}$:
\begin{align}
\boldsymbol{A}_{\bm{\rho}}&=\boldsymbol{P}_M \boldsymbol{\sigma}_{-} \boldsymbol{\sigma}_{+}^{-1}, \\
\boldsymbol{b}_{\bm{\rho}}&=\left(\boldsymbol{\sigma}_{+}^{-1} \boldsymbol{\mu}\right)^*=\boldsymbol{P}_M \boldsymbol{\sigma}_{+}^{-1} \boldsymbol{\mu},
\end{align}
where $\bm{\sigma}_{\pm} = \boldsymbol{\sigma} \pm \frac{1}{2} \bm{\mathbb{1}}_{2 M}$ and $\boldsymbol{P}_M=\left[\begin{array}{cc}
\bm{0}_M & \bm{\mathbb{1}}_M \label{eq:A_rho}\\
\bm{\mathbb{1}}_M & \bm{0}_M
\end{array}\right]$. \label{eq:b_rho}\\
If $\bm{\mathcal{G}}$ is a state vector $\bm{\psi}$, then $\bm{A}_{\bm{\psi}}$ and $\bm{b}_{\bm{\psi}}$ can be obtained from:
\begin{align}
\boldsymbol{A}_{\bm{\rho}} &=\boldsymbol{A}_{\bm{\psi}}^* \oplus \boldsymbol{A}_{\bm{\psi}}, \\
\boldsymbol{b}_{\bm{\rho}} &=\boldsymbol{b}_{\bm{\psi}}^* \oplus \boldsymbol{b}_{\bm{\psi}}.
\end{align}

\begin{figure}[!htb] 
\centering
\begin{subfigure}{0.9\linewidth}
  \includegraphics[width=\linewidth]{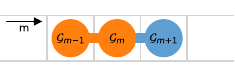}
  \vspace*{-2.5em}
  \caption{1-dimensional $\bm{\mathcal{G}}$\\
  (i.e.~state vector on 1 mode)}
  \label{fig:hypercross_dim1}
  \vspace*{0.5em}
\end{subfigure}
\begin{subfigure}{0.9\linewidth}
  \includegraphics[width=\linewidth]{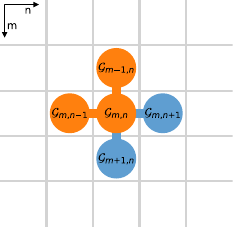}
  \caption{2-dimensional $\bm{\mathcal{G}}$\\
  (i.e.~state vector on 2 modes or density matrix on 1 mode)}
  \label{fig:hypercross_dim2}
\end{subfigure}
\caption{Schematic representation of how \cref{eq:recrel} can be used to calculate the Fock amplitudes $\mathcal{G}_{\bm{k}}$ of a Gaussian state. Every Fock index $\bm{k}$ marks a position in the Fock lattice. Every blue node can be written as a linear combination of the orange nodes.}
\label{fig:hypercross_dim1and2}
\end{figure}

Let us now define the `weight' of a Fock index $\bm{k}$ as:
\begin{equation}
    w = \sum_{i=1}^D k_i~.
\end{equation}
We see that \cref{eq:recrel} allows us to write $D$ Fock amplitudes of weight $w+1$ as linear combinations of a single Fock amplitude of weight $w$ and $D$ Fock amplitudes of weight $w-1$. In order to refer to these different roles, we call `read' the group of amplitudes of weight $w-1$ and `write' the group of amplitudes of weight $w+1$ (to refer to the fact that $D$ amplitudes need to be read from memory so that $D$ new ones can be written to memory), and we refer to the single amplitude of weight $w$ as the `pivot'.
\cref{fig:hypercross_dim1and2} gives a schematic representation of \cref{eq:recrel} for the case where $\bm{\mathcal{G}}$ is 1-dimensional (i.e.~for a state vector on one mode) and 2-dimensional (i.e.~for a state vector on two modes or a density matrix on one mode). In this figure, the amplitudes marked in blue (write) are written as linear combinations of the orange ones (read+pivot). In general, a Fock index $\bm{k}$ marks a position in a $D$-dimensional `Fock lattice'.
\cref{eq:recrel} can thus be interpreted as a relation between $2D+1$ amplitudes that we can draw as a cross (or hypercross for higher dimensions). We can repeatedly reposition the hypercross in $\bm{\mathcal{G}}$ to calculate new Fock amplitudes under the condition that we already computed the read and pivot amplitudes.

\subsection{State vector simulations}
\label{sec:state_vectors}

Let us now consider how we can apply the recurrence relation (that is, how we can move around the hypercross) to obtain the probabilities of PNR outcomes or the amplitudes of conditional states using the state vector formalism in a noiseless, lossless circuit. As the number of possible measurement results $\bm{n}=[n_1,n_2,...,n_M]$ ($n_i \in [0,1,...,\infty]$) is in principle infinite, we limit ourselves to calculating the most probable ones such that the required resources for our simulation remain finite.
We will consider the Fock amplitudes $\mathcal{G}_{\bm{k}}$ for all $\bm{k}$ of length $M$ that satisfy the following boundary conditions:
\begin{equation}
    \bm{0} \leq \bm{k} < \cutoffs~.
\label{eq:local_boundary_conditions}
\end{equation}
Here, $\cutoffs = [C_1,C_2,C_3,...]$ is the set of upper bounds for the photon numbers in all modes. We assume that they are chosen such that the probability of detecting $C_i$ or more photons in mode $i$ is negligible.

Note that Fock amplitude $\mathcal{G}_{\bm{0}}$ (where $\bm{0} = [0,0,...,0]$) is the vacuum component of $\bm{\mathcal{G}}$. If $\bm{\mathcal{G}}$ is a density matrix $\bm{\rho}$, it can be computed as:
\begin{equation}
    \rho_{\bm{0}} = \frac{\exp \left[-\frac{1}{2} \overline{\boldsymbol{\mu}}^{\dagger} \boldsymbol{\sigma}_{+}^{-1} \overline{\boldsymbol{\mu}}\right]}{\sqrt{\operatorname{det}\left(\boldsymbol{\sigma}_{+}\right)}},
\end{equation}
If $\bm{\mathcal{G}}$ is a state vector $\bm{\psi}$, ignoring a global phase, it holds that $\psi_{\bm{0}} = \sqrt{\rho_{\bm{0}}}$.

Starting from $\mathcal{G}_{\bm{0}}$, we can calculate all of the amplitudes by applying \cref{eq:recrel}.
We start by placing the pivot of our hypercross at $\bm{0}$ (for which $w=0$) and write amplitudes for which $w=1$. 
Next, we apply all pivots for which $w=1$ and write amplitudes for which $w=2$. By repeatedly increasing $w$ and applying all pivots of that weight, we can calculate the required amplitudes. 
As the amplitudes we write have a higher weight than the amplitudes we read, we know that the right amplitudes are always calculated before we need to read them. 

\begin{figure}[!htb]
\centering
\begin{subfigure}{.9\linewidth}
  \includegraphics[width=\linewidth]{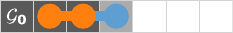}
  \caption{1 mode}
  \label{fig:state_vector_M1}
  \vspace*{0.4em}
\end{subfigure}
\begin{subfigure}{.9\linewidth}
  \includegraphics[width=\linewidth]{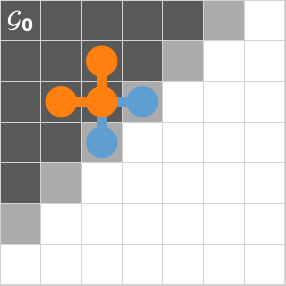}
  \caption{2 modes}
  \label{fig:state_vector_M2}
\end{subfigure}
\caption{Intermediate step of state vector simulations for circuits consisting of 1 and 2 modes. 
Fock amplitudes $\mathcal{G}_{\bm{k}}$ of the output state vectors are computed recursively. We start at $\mathcal{G}_{\bm{0}}$ and apply pivots in order of increasing weight until all amplitudes are calculated. At this intermediate step, dark grey cells have been used as pivots. Light grey cells have been written and will be used as pivots in the next step.
Animated versions of these figures are included in the Supplementary Materials.
}
\label{fig:state_vector}
\end{figure}

\cref{fig:state_vector} shows an intermediate step of this process for circuits that consist of one and two modes. In this figure, the cutoff values of all modes are chosen to be 7. Dark grey cells depict amplitudes that have already been used as pivots. Light grey cells are amplitudes that have been calculated, but have not yet been used as pivots. At the end of the process all cells in the figure will be calculated.

Note that this strategy to calculate Fock amplitudes allows for two types of parallelization.
First, given a specific pivot, we can parallelize the calculations of different elements in the `write' group. 
Second, since we order pivots according to increasing weight, we can also apply pivots of the same weight simultaneously.

\subsection{Alternative cutoff conditions}

The boundary conditions of \cref{eq:local_boundary_conditions} are useful for simulating circuits for which we know the maximum number of photons that a PNR detectors can measure. 
The cutoff in the undetected modes can be chosen separately, depending on the required accuracy for calculating the conditional state.
However, the recurrence relation also allows one to consider other cutoff conditions.

A first useful example occurs when we want to place an upper bound on the total number of photons that is present in all modes. As the total number operator $ \bm{\hat{n}} = \sum_{i=1}^M \hat{n}_i$ commutes with the multi-mode Fock Hamiltonian \cite{gerry2005introductory}, such an upper bound defines a cutoff on the energy levels of the multi-mode Gaussian state before the detectors. More formally, we can replace \cref{eq:local_boundary_conditions} by:
\begin{equation}
    0 \leq w(\bm{k}) < w_\text{max}~,
\label{eq:global_boundary_conditions}
\end{equation}
which can be related to an upper bound for the total number of photons $N_\text{max}$ in the circuit: 
\begin{equation}
    w_\text{max} =
    \begin{cases}
        N_\text{max}, &         \text{if $\bm{\mathcal{G}}$ is a state vector},\\
        2 N_\text{max}, &         \text{if $\bm{\mathcal{G}}$ is a density matrix.}
    \end{cases}
\end{equation}

Note that for \cref{eq:global_boundary_conditions} the number of amplitudes $\mathcal{G}_{\bm{k}}$ that have the same weight increases binomially with $w$.
For \cref{eq:local_boundary_conditions}, this number of amplitudes first increases with $w$, after which it reaches a maximum and decreases. Indeed, once $w \geq \text{min}(\cutoffs)$, the right inequality of \cref{eq:local_boundary_conditions} starts to exclude general Fock indices of weight $w$. Eventually, when $w$ is raised all the way to $\sum_{i=1}^D (C_i-1)$ the number of allowed indices has decreased back to 1.

Another possible cutoff condition is given by the total sum of the probabilities of PNR outcomes. After each iteration (in which we apply all pivots of weight $w$), we can evaluate this sum and check whether it is sufficiently close to 1 to stop the process.

\subsection{Circuits without displacement gates}
\label{sec: no displacement intro)}

In Reference \cite{yao2022recursive} we showed how to compute the parameters $\bm{A}$, $\bm{b}$ and $\mathcal{G}_{\bm{0}}$ that define a Gaussian operator. More specifically for Gaussian states, we showed how $\bm{A}$, $\bm{b}$ and $\mathcal{G}_{\bm{0}}$ can be calculated from the covariance matrix and means vector. Moreover, it can be shown that for a state with zero displacement vector we have $\bm{b} = \bm{0}$. Note that this applies to the states before the detectors in \cref{fig:example_circuits} as these circuits do not contain displacement gates.

In the case that there is no displacement, we can substitute $\bm{b} = \bm{0}$ in \cref{eq:recrel} such that our recurrence relation turns into:
\begin{equation}
    \mathcal{G}_{\bm{k} + \bm{1}_i} = \frac{1}{\sqrt{k_i+1}} \sum_{l = 1}^{D}\sqrt{k_l} \: \mathcal{G}_{\bm{k} - \bm{1}_l}A_{il}~.\label{eq:recrel_noB}
\end{equation}

\begin{figure}[!htb]
\vspace*{0.8em}
\centering
\begin{subfigure}{.9\linewidth}
  \includegraphics[width=\linewidth]{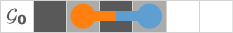}
  \caption{1 mode}
  \label{fig:state_vector_M1_noDisp}
  \vspace*{0.4em}
\end{subfigure}
\begin{subfigure}{.9\linewidth}
  \includegraphics[width=\linewidth]{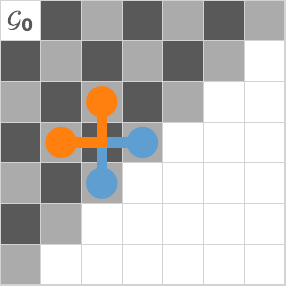}
  \caption{2 modes}
  \label{fig:state_vector_M2_noDisp}
\end{subfigure}
\caption{
Intermediate step of state vector simulations for circuits that do not contain displacement gates (consisting of 1 and 2 modes).
Fock amplitudes $\mathcal{G}_{\bm{k}}$ are calculated recursively as in \cref{fig:state_vector}, but now  they are zero when $\sum_i k_i$ is odd. Consequently, pivots (i.e.~the central nodes of the hypercross) do not need to be read.
At this intermediate step, dark grey cells have been used as pivots but only for placing the cross (their value remains zero), while light grey cells have been actually written.
}
\label{fig:state_vector_noDisp}
\end{figure}

We find that the only Fock amplitudes that differ from zero are the ones which have a Fock index $\bm{k}$ with even weight. For state vectors, we can alter the strategy described in \cref{fig:state_vector} by only considering pivots that have odd weight. This leads to the checkered pattern of \cref{fig:state_vector_noDisp}, where we still apply pivots in order of increasing weight. 
Note that now we now fill the array twice as fast because we only need to compute half of the amplitudes.

\subsection{Gradients}
\label{sec:gradients}
In this section, we present how the framework above allows not only to \textit{simulate} but also to \textit{optimize} circuits. Given a loss function $L$ that depends on the probabilities of the PNR outcomes (and the conditionally generated states), we need to calculate the partial derivatives of $L$ with respect to the parameters of the circuit. 
As explained in Reference \cite{miatto2020fast}, the so-called `down-stream gradient' of $L$ with respect to the conjugate of a complex circuit parameter $\xi$ can be computed using the chain rule as follows:
\begin{equation}
    \frac{\partial L}{\partial \xi^*}=\sum_{\boldsymbol{k}} \frac{\partial L}{\partial \mathcal{G}_{\boldsymbol{k}}^*} \frac{\partial \mathcal{G}_{\boldsymbol{k}}^*}{\partial \xi^*}+\frac{\partial L}{\partial \mathcal{G}_{\boldsymbol{k}}} \frac{\partial \mathcal{G}_{\boldsymbol{k}}}{\partial \xi^*}~.
\end{equation}

We now consider $\xi$ to be equal to $b_m$ or $A_{mn}$ and note that \cref{eq:recrel} does not depend on $b_m^*$ or $A_{mn}^*$, such that:

\begin{gather}
    \frac{\partial L}{\partial b_{m}^*}=\sum_{\boldsymbol{k}} \frac{\partial L}{\partial \mathcal{G}_{\boldsymbol{k}}^*} \frac{\partial \mathcal{G}_{\boldsymbol{k}}^*}{\partial b_{m}^*}=\sum_{\boldsymbol{k}} \frac{\partial L}{\partial \mathcal{G}_{\boldsymbol{k}}^*} \left(\frac{\partial \mathcal{G}_{\boldsymbol{k}}}{\partial b_{m}}\right)^* ~,\label{eq:dL_dB}\\
    \frac{\partial L}{\partial A_{mn}^*}=\sum_{\boldsymbol{k}} \frac{\partial L}{\partial \mathcal{G}_{\boldsymbol{k}}^*} \frac{\partial \mathcal{G}_{\boldsymbol{k}}^*}{\partial A_{mn}^*}=\sum_{\boldsymbol{k}} \frac{\partial L}{\partial \mathcal{G}_{\boldsymbol{k}}^*} \left(\frac{\partial \mathcal{G}_{\boldsymbol{k}}}{\partial A_{mn}}\right)^*~.\label{eq:dL_dA}
\end{gather}

As the upstream gradient tensor $\partial L / \partial \mathcal{G}_{\bm{k}}^*$ can be provided to us by an automatic differentiation framework such as TensorFlow or PyTorch, we only have to compute the local gradients $\partial \mathcal{G}_{\bm{k}} / \partial b_m$ and $\partial \mathcal{G}_{\bm{k}} / \partial A_{mn}$.

From \cref{eq:recrel} we now derive:

\begin{widetext}
\begin{gather}
\frac{\partial \mathcal{G}_{\bm{k} + \bm{1}_i}}{\partial b_m} = \frac{1}{\sqrt{k_i+1}}\left(\frac{\partial \mathcal{G}_{\bm{k}}}{\partial b_m}b_i + \mathcal{G}_{\bm {k}} \delta_{im} + \sum_{l = 1}^{D}\sqrt{k_l} \: \frac{\partial \mathcal{G}_{\bm{k} - \bm{1}_l}}{\partial b_m}A_{il}\right) \label{eq:recrel_dB} ~,\\
\frac{\partial \mathcal{G}_{\bm{k} + \bm{1}_i}}{\partial A_{mn}} = \frac{1}{\sqrt{k_i+1}}\left(\frac{\partial \mathcal{G}_{\bm{k}}}{\partial A_{mn}}b_i + \sum_{l = 1}^{D}\sqrt{k_l}\left[\frac{\partial \mathcal{G}_{\bm{k} - \bm{1}_l}}{\partial A_{mn}}A_{il} + \mathcal{G}_{\bm{k} - \bm{1}_l} \delta_{im}\delta_{ln}\right]\right)~, \label{eq:recrel_dA}
\end{gather}
\end{widetext}%
where $\delta_{jk}$ is the Kronecker delta function. 
Since both \cref{eq:recrel_dB} and \cref{eq:recrel_dA} are structured in a similar way as \cref{eq:recrel}, we can implement all three equations simultaneously. We do so by taking a single walk through the Fock lattice, that is, by performing a single iteration over the Fock indices $\bm{k}$.
We still differentiate between the different types of Fock indices `read' ($\bm{k}-\bm{1}_l$), `pivot' ($\bm{k}$) and `write' ($\bm{k}+\bm{1}_i$), but instead of only manipulating amplitudes $\mathcal{G}_{\bm{k}}$, we now also process their partial derivatives with respect to $b_m$ and $A_{mn}$. 
Note that every $\bm{k}$ now corresponds with one Fock amplitude $\mathcal{G}_{\bm{k}}$, $D$ gradients $\partial \mathcal{G}_{\bm{k}} /\partial b_m$ and $D^2$ gradients $\partial \mathcal{G}_{\bm{k}} / \partial A_{mn}$, such that both the memory and time usage of an \textit{optimization} are a factor $1+D+D^2$ higher than those of a \textit{simulation}.

\section{Extension to density matrix simulations}
\label{sec:algos density matrices}
\subsection{Algorithm for Gaussian Boson Sampling} %Detecting all modes
\label{sec: diagonal algo}
Consider a circuit of which all $M$ modes are detected (such as the one in \cref{fig:circuit_detectAllModes}).
To capture mixed states (such as can arise in the presence of photon loss) density matrices must be used in place of state vectors.
For simplicity, let us assume that the photon number cutoff in each mode is equal to $C$. To calculate the probabilities of the $C^M$ possible PNR detection patterns, one could start by following the procedure described in \cref{sec:state_vectors} to calculate all $C^{2M}$ Fock amplitudes of the multi-mode state before the detector. The probability of observing a certain photon number pattern $\bm{n}=[n_1,n_2,...,n_M]$ at the detectors is then given by:
\begin{equation}
    p(\bm{n}) = \mathcal{G}_{n_1 n_1 n_2 n_2 ... n_M n_M}~.
\end{equation}
However, as we are only interested in the $C^M$ diagonal amplitudes, we can construct a more efficient algorithm that selectively applies the recurrence relation in the Fock lattice. This way we prevent the calculation of irrelevant amplitudes as much as possible. After choosing an adequate set of pivot positions, we can apply them in order of increasing weight. 

\subsubsection{Single mode}
\label{sec: single mode}

\begin{figure*}[!htb]
\centering
\begin{subfigure}{.45\linewidth}
  \includegraphics[width=\linewidth]{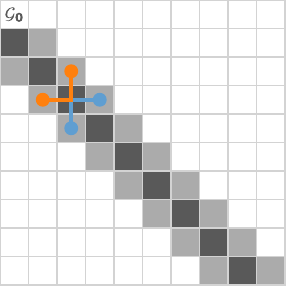}
  \caption{Circuit without displacement gates}
  \label{fig:detectAllModes_M1_C20_withoutDisplacement}
\end{subfigure}
\hspace*{\fill}%
\begin{subfigure}{.45\linewidth}
  \includegraphics[width=\linewidth]{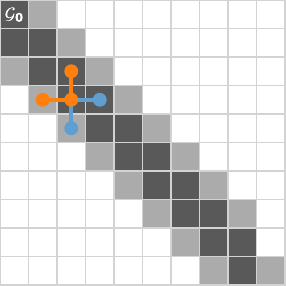}
  \caption{Circuit with displacement gates}
  \label{fig:detectAllModes_M1_C20}
\end{subfigure}
\caption{
Visualisation of how \cref{eq:recrel} can be applied to density matrices in order to calculate the detection probabilities $|\mathcal{G}_{n_1 n_1}|^2$ of a single mode circuit.
Pivots (dark grey) are applied from top left to bottom right, i.e.~in order of increasing weight. Light grey cells are non-pivot amplitudes that are written. White cells do not have to be written, which improves on the naive idea of applying pivots in all cells (as in \cref{fig:state_vector_M2}). In Fig.~a, the pivots are not read as there \cref{eq:recrel} simplifies to \cref{eq:recrel_noB}.
We chose to upper bound the photon number by 10 in this example.
Animated versions of these figures are included in the Supplementary Materials.
}
\label{fig:nestedRepr_detectAllModes_M1}
\end{figure*}

Let us first consider the case where we have a single mode. Here the Fock lattice only has two dimensions (i.e.~$\bm{k}=[m,n]$) and we can use the hypercross of \cref{fig:hypercross_dim2}.
For now also consider the case where the circuit under consideration does not contain displacement gates. As explained in \cref{sec: no displacement intro)}, this implies that the inner `pivot' node of the hypercross cross does not need to be read.
\cref{fig:detectAllModes_M1_C20_withoutDisplacement} visualizes how  \cref{eq:recrel_noB} can be applied in order to calculate the required diagonal amplitudes.
We have chosen all pivots of the type $[a+1,a]$ that satisfy $[0,0]\leq[a+1,a]<[C_1,C_1]$. Note that we could have equivalently chosen pivots of the type $[a,a+1]$ instead. We apply the pivots in order of increasing weight, i.e.~from the top left to the bottom right.
As these pivots only read amplitudes that are previously written by other pivots, the total set of pivots can be said to be `self-sufficient'.

\cref{fig:detectAllModes_M1_C20} shows the case where the circuit under consideration does contain displacement gates.
Now, we also have to read the value of the pivot node in order to apply the hypercross.
These values (at positions $[a+1,a]$) can be provided by introducing extra pivots of the type $[a,a]$. 
In their turn, the off-diagonal pivots provide the amplitude values of the diagonal pivots.
In other words, the total set of the diagonal and off-diagonal pivots is self-sufficient here.

\subsubsection{Two modes}
\begin{figure*}[!htb]
\centering
  \includegraphics[width=0.725\linewidth]{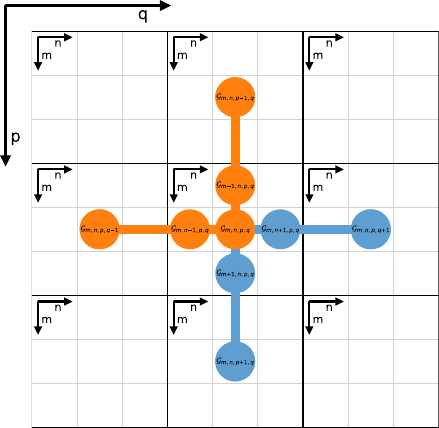}\hspace*{2.2em}
\caption{Schematic representation of \cref{eq:recrel} where $\bm{\mathcal{G}}$ is 4-dimensional. The Fock amplitudes $\bm{\mathcal{G}}_{mnpq}$ are represented via the Kronecker product:
all $C_1 \times C_1$ corresponding to different values of $p$ and $q$ are combined in a block matrix.
}
\label{fig:hypercross_M2}
\end{figure*}

\begin{figure}[!htb]
\centering
  \includegraphics[width=1\linewidth]{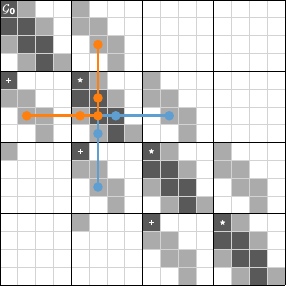}
\caption{Visualisation of how \cref{eq:recrel} (i.e.~the hypercross from \cref{fig:hypercross_M2}) can be applied to density matrices in order to calculate the detection probabilities $\mathcal{G}_{n_1 n_1 n_2 n_2}$ of a two mode circuit.
The photon number in both modes is upper bound by 4.
Dark grey cells represent pivots. Light grey cells represent non-pivot amplitudes that are written. 
The pivots marked as $\Plus$ write to the pivots marked as $\bm{\star}$. Similar to $\mathcal{G}_{\bm{0}}$, these last pivots ($\bm{\star}$) act as `seed' amplitudes in their $C_1 \times C_1$ blocks. 
An animated version of this figure is included in the Supplementary Materials.}
\label{fig:nestedRepr_detectAllModes_M2}
\end{figure}

We now consider density matrix simulations of GBS circuits with two modes, such that \cref{eq:recrel} can be represented by a four dimensional hypercross. However, we still choose to visualize both the hypercross and $\bm{\mathcal{G}}$ in two dimensions via the Kronecker product.
Below, we explain in more detail how such a representation is constructed. The hypercross itself is shown in \cref{fig:hypercross_M2}. \cref{fig:nestedRepr_detectAllModes_M2} visualizes how this hypercross can be applied to get the diagonal Fock amplitudes in the case where $\cutoffs = [4,4]$.

We write $\bm{k} = [m,n ,p,q]$, where $[m,n]$ and $[p,q]$ are the indices corresponding to the first and second mode respectively. Note now that if $[p,q]$ would be fixed, we are left with a 2D matrix that is only indexed by $[m,n]$, such that it can be visualized in a similar way as \cref{fig:nestedRepr_detectAllModes_M1}. We now combine all such matrices (for all possible values of $p$ and $q$) in a block matrix. This leads to a 2D `nested representation'. If $M>2$, we can recursively apply this process for different index pairs (i.e.~constructing block matrices of block matrices), such that we always end up with a 2D image. 
Note that pivots are no longer applied from top left to bottom right in this representation, as this would not correspond with the order of increasing weight.

% some insight that connects with the section on a single mode
We have to make sure that amplitudes are written before they are read. In other words, the total set of pivots used in \cref{fig:nestedRepr_detectAllModes_M2} has to be self-sufficient. We can check that this is true by first considering the pivots of the type $[a,a,b,b]$ and $[a+1,a,b,b]$ (i.e.~the diagonal cells in \cref{fig:nestedRepr_detectAllModes_M2} and the cells under those). This set of pivots is \textit{almost} self-sufficient: within each $C_1 \times C_1$ block that lies on the diagonal of \cref{fig:nestedRepr_detectAllModes_M2} (i.e.~within each block containing amplitudes of the type $[m,n,b,b]$), \textit{almost} all of these pivots get their required `read' and `pivot' amplitudes from the `write' amplitudes from another pivot in those blocks. The only amplitudes that are missing to complete the self-sufficiency are the amplitudes of the type $[0,0,b,b]$ (marked as $\bm{\star}$). These last amplitudes act like `seed amplitudes' in the diagonal $C_1 \times C_1$ blocks, similar to how $\mathcal{G}_{\bm{0}}$ acts as a seed in \cref{fig:nestedRepr_detectAllModes_M1}. These missing amplitudes can be obtained from the remaining pivots outside of the diagonal $C_1 \times C_1$ blocks: $[0,0,b{+}1,b]$ (marked as $\Plus$). 
These last pivots `bridge' the gaps between different diagonal $C_1 \times C_1$ blocks by providing the necessary increments of $k_i$ for $i \in \{3,4,5,...,2M\}$.

\subsubsection{General number of modes}
\label{sec: algo1}

The pivot placement strategy of \cref{fig:nestedRepr_detectAllModes_M1,fig:nestedRepr_detectAllModes_M2} can be generalized to a larger number of modes. The strategy for 3 modes is visualized in \cref{sec:nestedRepr_detectAllModes_M3}. 

\cref{algo: detectAllModes pivotPlacement} shows how a GBS circuit with an \textit{arbitrary} number of modes can be simulated in the density matrix formalism. 
Lines \ref{line 1} to \ref{line 5} are used to apply the diagonal pivots $diag=[a,a,b,b,c,c,...]$ in order of increasing weight. Note again that these pivots are also diagonal in the nested representation, while the order in which we apply them is not necessarily from top left to bottom right (see for example the animated version of \cref{fig:nestedRepr_detectAllModes_M2} in the Supplementary Materials). In order to apply the diagonal pivots, a variable $\sumvar$ is increased stepwise, starting from 0. Each time, we apply all diagonal pivots that satisfy both $a+b+c+...=\sumvar$ and the boundary conditions of \cref{eq:local_boundary_conditions}.

Lines \ref{line 6} to \ref{line 10} are used to apply the off-diagonal pivots $\diag + \bm{1}_{2\dvar-1}$, where $\dvar \in \{1,2,...,M\}$ (i.e.~$[a{+}1,a,b,b,c,c,...]$, $[a,a,b{+}1,b,c,c,...]$, $[a,a,b,b,c{+}1,c,...]$, etc.). For $\dvar{=}1$, the off-diagonal pivots lie in the diagonal $C_1 \times C_1$ blocks. For $\dvar>1$, the off-diagonal pivots are `bridge pivots' that provide the `source amplitudes' $[0,0,b,b,c,c,...]$. Note that because of line \ref{line 8}, the number of off-diagonal pivots decreases with $\dvar$ (see both \cref{fig:nestedRepr_detectAllModes_M2,sec:nestedRepr_detectAllModes_M3} for reference).

\begin{algorithm*}[!htb]
\caption{Density matrix simulation of a GBS circuit}
\label{algo: detectAllModes pivotPlacement}
\begin{algorithmic}[1]
\For{$\sumvar \gets 0$ to $(\sum_{i=1}^{M}C_i)-1$} \label{line 1}
    \Comment{Stepwise increase of pivot weight}
    \State calculate the set $diag\_set$ of all length-$2M$ indices $[a,a,b,b,c,c,...]$
    \Statex[1] that satisfy $a+b+c+...=\sumvar$ and $\bm{0} \leq [a,b,c,...] < \cutoffs$ \label{line 2}
    \For{$diag$ in $diag\_set$}
        \If{$diag_1<C_1-1$} \label{line 4}% prevents diag pivot being used of the type [C-1,C-1,b,b,c,c,...]
            \State \textbf{apply} $diag$ as pivot \Comment{Diagonal pivot ($w=2 \sumvar$)} \label{line 5}
        \EndIf
    \EndFor
    \For{$diag$ in $diag\_set$} \label{line 6}
        \For{$\dvar \gets 1$ to $M$}
            \If{the first $2(\dvar-1)$ elements of $diag$ are 0} \label{line 8}
                \If{$diag_{2\dvar}<C_{\dvar}-1$} \label{line 9}% ensures that pivot is below cutoff
                    \State \textbf{apply} $\diag + \bm{1}_{2\dvar-1}$ as pivot \Comment{Off-diagonal pivot ($w=2 \sumvar + 1$)} \label{line 10}
                \EndIf
            \EndIf
        \EndFor
    \EndFor
\EndFor
\end{algorithmic}
\end{algorithm*}

In \cref{sec:analysis algo1}, we show that both the total number of pivots and the total number of written amplitudes that appear in \cref{algo: detectAllModes pivotPlacement} scale like $\prod_{i=1}^M C_i$, which simplifies to $C^M$ if the cutoffs on all modes are equal.

Note that if the local cutoff conditions of \cref{eq:local_boundary_conditions} are replaced by the global cutoff condition of \cref{eq:global_boundary_conditions}, then the sum of line \ref{line 1} runs to $N_\text{max} = \frac{1}{2} w_\text{max}$ instead, while the cutoff conditions in lines \ref{line 2}, \ref{line 4} and \ref{line 8} drop out. As shown in \cref{sec:analysis algo1}, the scaling of the algorithm then changes to $(w_\text{max})^M$.

\subsubsection{Compact storage of the Fock amplitudes}
\label{sec: buffer strategy}
In \cref{sec:analysis algo1}, we show that all amplitudes that are written in \cref{algo: detectAllModes pivotPlacement} can be parameterized as $\diag + \offset$ where $\diag$ is a diagonal position in the Fock lattice and $\offset$ is an offset vector that only comes in a select number of types.
This parametrization helps to store the amplitudes in a unique and compact manner. However, in the case that we detect all modes, we are only interested in the $\prod_{i=1}^M C_i$ diagonal amplitudes. The off-diagonal amplitudes do not need long-term storage in memory.
It can be shown that all off-diagonal amplitudes are included in the `read' group of a pivot exactly once. 
Thus, we can remove off-diagonal 'read' amplitudes from memory once they have been used.
We only have to store a buffer of off-diagonal amplitudes that correspond with a select number of weight values. 
In addition to the animated versions of \cref{fig:nestedRepr_detectAllModes_M1,fig:nestedRepr_detectAllModes_M2,fig:nestedRepr_detectAllModes_M3}, we also include animations in the Supplementary Materials that apply this `buffer strategy'.

\begin{figure}[!htb]
    \centering
    \includegraphics[width=1\linewidth]{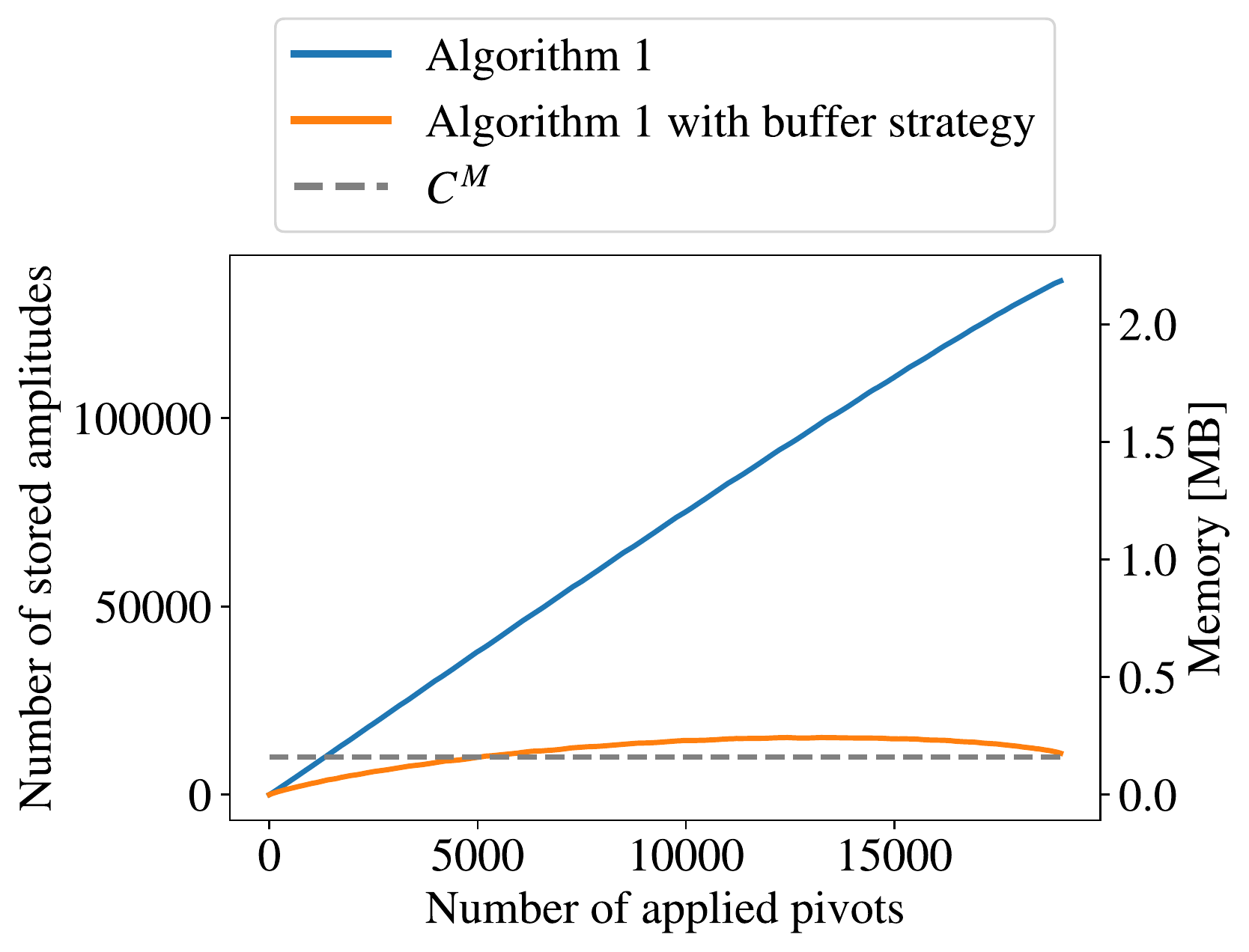}
    \caption{The number of stored amplitudes when using a buffer of off-diagonal amplitudes (orange curve) and without using such a buffer (blue curve) as a function of the number of applied pivots. After all pivots are applied, the buffer is left empty such that the orange curve reaches a value of $C^M$ (dashed curve).
    This figure is made using 4 modes, all with photon number cutoff of 10. Both the real and complex part of the amplitudes are stored as 64-bit-precision floating-point numbers. % np.complex128
    }
    \label{fig:detectAllModes_M4_bufferOverTime}
\end{figure}

For a circuit consisting of 4 modes (such as the one in \cref{fig:circuit_detectAllModes}), \cref{fig:detectAllModes_M4_bufferOverTime} shows how the number of stored amplitudes evolves as we apply more pivots. We have chosen the photon number cutoff to be 10 in all modes. In contrast to the strategy without buffer (blue curve), the buffer strategy (orange curve) reaches a maximum before the end of the algorithm is reached. This results from the fact that the number of pivots that have an equal weight reaches a maximum at $w = \sum_{i=1}^M C_i$ when we apply the local boundary conditions of \cref{eq:local_boundary_conditions}.
For reference, \cref{fig:detectAllModes_M4_bufferOverTime} also shows a horizontal dashed line at $\prod_{i=1}^M C_i=C^M=10^4$. Note that after completing \cref{algo: detectAllModes pivotPlacement} using the buffer strategy all off-diagonal amplitudes are removed, such that the orange curve coincides with the dashed curve.

\subsection{Algorithm for conditional state generation} %Detecting all but one mode
\label{sec: leftover algo}

Let us now consider circuits where all but one mode are detected, such as the one of \cref{fig:circuit_leftovermode}. Our results can readily be generalized to an arbitrary number of undetected modes.
Our goal is now to calculate the distribution of states that are generated conditionally on the PNR detection results. 
As a first example, we consider a circuit with two modes and one detector, such that we can use the nested representation of \cref{fig:hypercross_M2}.
In this representation, the targeted distribution is defined by the Fock amplitudes $\mathcal{G}_{mnpq}$ in the diagonal $C_1 \times C_1$ blocks. Each detection outcome corresponds with one such $C_1 \times C_1$ block, which is the unnormalized density matrix of the conditional state. 

\begin{figure*}[!htb]
\centering
\begin{subfigure}{.45\linewidth}
  \centering
  \includegraphics[width=\linewidth]{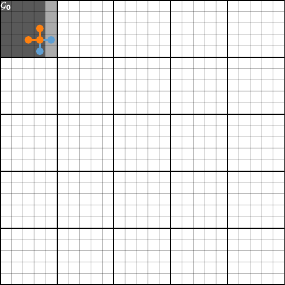}
  \caption{Calculate upper left $C_1 \times C_1$ block}
  \label{fig:nestedRepr_leftoverMode_M2_subfigB}
\end{subfigure}
\hspace*{\fill}%
\begin{subfigure}{.45\linewidth}
  \centering
  \includegraphics[width=\linewidth]{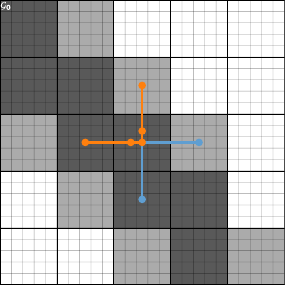}
  \caption{Apply coarse-grained version of \cref{algo: detectAllModes pivotPlacement}}
  \label{fig:nestedRepr_leftoverMode_M2_subfigA}
\end{subfigure}
\caption{
Visualisation of \cref{algo: leftoverMode pivotPlacement} for a circuit of two modes, one of which is detected.
Dark grey cells represent pivots. Light grey cells represent non-pivot amplitudes that are written. White cells represent amplitudes that are not written. For the pivots $\mathcal{G}_{3300}$ (in Fig.~a) and $\mathcal{G}_{2222}$ (in Fig.~b) the hypercross is shown, where we only keep two of its blue nodes.
For both modes, we chose a photon number cutoff of 5.
Animated versions of these figures are included in the Supplementary Materials.}
\label{fig:nestedRepr_leftoverMode_M2}
\end{figure*}

The targeted blocks can be calculated using the two step process presented in \cref{fig:nestedRepr_leftoverMode_M2}. 
First, we calculate all Fock amplitudes in the upper left $C_1 \times C_1$ block, which is the density matrix corresponding with detecting zero photons. For this first step, we can use the hypercross of \cref{fig:hypercross_M2} where we choose only to \textit{increment} indices $m$ and $n$ (not $p$ and $q$). Note that we also do not have to \textit{decrement} $p$ and $q$, as these amplitudes would correspond with negative photon numbers. 
For the second step of our simulation process, we do have to \textit{decrement} all indices, but this time we choose only to \textit{increment} indices $p$ and $q$ (not $m$ and $n$). Moreover, we choose to apply pivots in blocks of size $C_1 \times C_1$. By doing so, we can apply a coarse-grained version of \cref{algo: detectAllModes pivotPlacement} \textit{as if} the circuit under consideration has $M-1$ modes. In this example, $M=2$ such that we apply a coarse grained version of \cref{fig:detectAllModes_M1_C20}. Within each $C_1 \times C_1$ block of pivots, the individual pivots still need to be applied according to increasing weight, similar to \cref{fig:state_vector_M2}.

This simulation process for state generator circuits can be generalized to an arbitrary number of modes $M$. 
\cref{algo: leftoverMode pivotPlacement} considers all cases where we have 1 undetected mode and $M-1$ detected modes. The extension to an arbitrary number of undetected modes is straightforward.
A similar two step process is followed as in \cref{fig:nestedRepr_leftoverMode_M2}.
Note that step 2 of \cref{algo: leftoverMode pivotPlacement} is indeed a coarse-grained version of \cref{algo: detectAllModes pivotPlacement} as we apply $C_1 \times C_1$ blocks of pivots. That is, we apply pivots $[m,n,a,a,b,b,c,c,...]$ for $m,n \in \{0,1,...,C_1-1\}$ where $a,b,c,...$ follow from \cref{algo: detectAllModes pivotPlacement} after substituting $M$ by $M-1$ and $\cutoffs$ by $[C_2,C_3,C_4,...]$.
As \cref{algo: detectAllModes pivotPlacement} scales as $\prod_{i=1}^M C_i$, it is clear from the above that \cref{algo: leftoverMode pivotPlacement} scales as ${C_1}^2 \prod_{i=2}^M C_i$. In the case where we choose all modes to have the same cutoff $C$, these scaling factors are $C^M$ and $C^{M+1}$ respectively.

\begin{algorithm*}[!htb]
\caption{Density matrix simulation of a conditional state generator circuit}
\label{algo: leftoverMode pivotPlacement}
\begin{algorithmic}[1]
\vspace*{0.5\baselineskip}%
\Algphase{\textbf{Step 1} Calculate $C_1 \times C_1$ block that corresponds with zero photon detections}
\State \textbf{apply} all pivots  $[m,n,0,0,...]$ for $m\in \{0,1,...,C_1-1\}$, $n\in \{0,1,...,C_1-2\}$ \label{algoLine:fillFirstBlock}
\Statex[0] \(\triangleright\) \textbf{only write} amplitude types $[m+1,n,0,0,...]$ and $[m,n+1,0,0,...]$
\Algphase{\textbf{Step 2} Coarse-grained version of \cref{algo: detectAllModes pivotPlacement}}
\For{$\sumvar \gets $0 to $(\sum_{i=2}^{M}C_i$)-1} % Note that i starts from 2 instead of 1.
    \State calculate the set $diag\_set$ of all indices $[a,a,b,b,c,c,...]$ of length $2(M-1)$ \label{algoLine: diagSet}
    \Statex[1] that satisfy $a+b+c+...=\sumvar$ and $\bm{0} \leq [a,b,c,...]<[C_2,C_3,C_4,...]$
    \For{$diag$ in $diag\_set$}
        \If{$diag_2<C_2-1$}
            \State \textbf{apply} all pivots $[m,n,diag]$ for $m,n \in \{0,1,...,C_1-1\}$ \label{algoLine:readWriteBlock1}
            \Comment{Diagonal $C_1 \times C_1$ block}
            \Statex[3] \(\triangleright\) \textbf{do not write} amplitude types $[m+1,n,0,0,...]$ and $[m,n+1,0,0,...]$
        \EndIf
    \EndFor
    \For{$diag$ in $diag\_set$}
        \For{$\dvar \gets 1$ to $M-1$}
            \If{the first $2(\dvar-1)$ elements of $diag$ are 0}
                \If{$diag_{2\dvar}<C_{1+\dvar}-1$}
                    \State \textbf{apply} all pivots $[m,n,diag] + \bm{1}_{2\dvar+1}$ for $m,n \in \{0,1,...,C_1-1\}$ \label{algoLine:readWriteBlock2} 
                    \Statex[5] \Comment{Off-diagonal $C_1 \times C_1$ block}
                    \Statex[5] \(\triangleright\) \textbf{do not write} amplitude types $[m+1,n,0,0,...]$ and $[m,n+1,0,0,...]$
                \EndIf
            \EndIf
        \EndFor
    \EndFor
\EndFor
\end{algorithmic}
\end{algorithm*}

\section{Complexity}
\label{sec:complexity}

In the case where we use state vectors, \cref{sec:state_vectors} explains how pivots can be applied to calculate all Fock amplitudes that satisfy the cutoff conditions of \cref{eq:local_boundary_conditions}. The total number of pivots then scales as $O(\prod_{i=1}^M C_i)$.
In the case where we simulate a GBS circuit using density matrices, we apply \cref{algo: detectAllModes pivotPlacement}. In \cref{sec:analysis algo1}, we show that the total number of pivots that are used in this algorithm also scales as $O(\prod_{i=1}^M C_i)$. 

As is clear from \cref{eq:recrel}, the complexity of applying a single pivot is given by $D^2$. (Note that \cref{eq:recrel} can be rewritten as the sum of a vector and a matrix-vector multiplication by rescaling $\mathcal{G}_{\bm{k} - \bm{1}_l}$ and $\mathcal{G}_{\bm{k} + \bm{1}_i}$ with $\sqrt{k_l}$ and $\sqrt{k_i+1}$ respectively.) From \cref{eq:cases_D} it follows that both using state vectors and density matrices, our algorithms for GBS simulation scale like $O(M^2 \prod_{i=1}^M C_i)$.
As is clear from \cref{sec: leftover algo}, for the generation of single mode conditional states, this complexity changes to $O(M^2 C_1^2 \prod_{i=2}^M C_i)$. \cref{algo: leftoverMode pivotPlacement} can readily be extended to account for a general number of undetected modes. By doing so, the complexity changes to:
\begin{equation}
    O(M^2 \prod_{i \in I_U} C_i^2 \prod_{i \in I_D} C_i),
\end{equation}
where $I_U$ and $I_D$ are the sets of indices $i$ that respectively correspond to undetected and detected modes.

In the remainder of this work, we first demonstrate how this scaling behaviour can be observed for circuits with 4 modes. Afterwards, the results for GBS circuits are compared to the state-of-the-art classical simulation method.

\subsection{Memory usage and simulation time}
\label{sec:space_and_time}
\begin{figure*}[!htb]
    \centering
    \begin{subfigure}{0.48\linewidth}
      \centering
      \includegraphics[width=\linewidth]{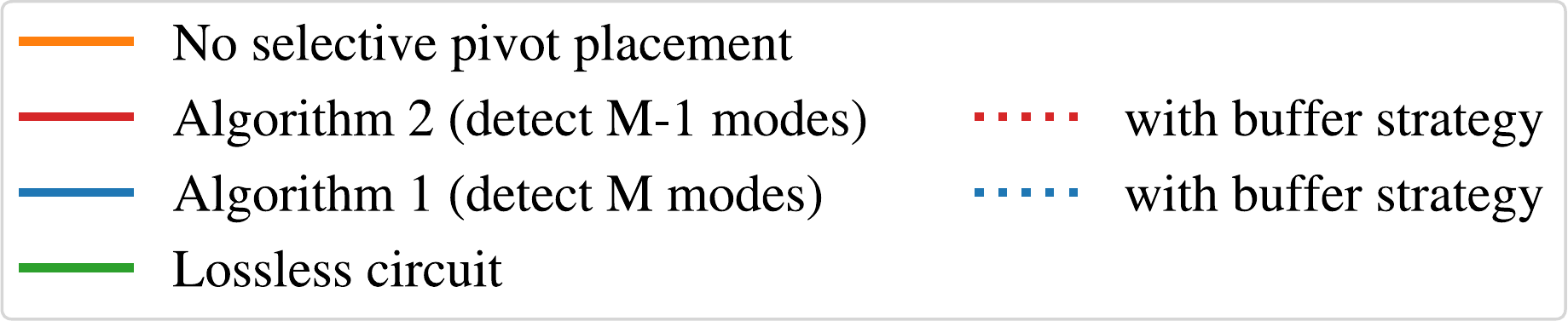}
    \end{subfigure}
    % blank line for new row
    \vspace*{-.8em}
    
    \begin{subfigure}{.49\linewidth}
      \centering
      \includegraphics[width=\linewidth]{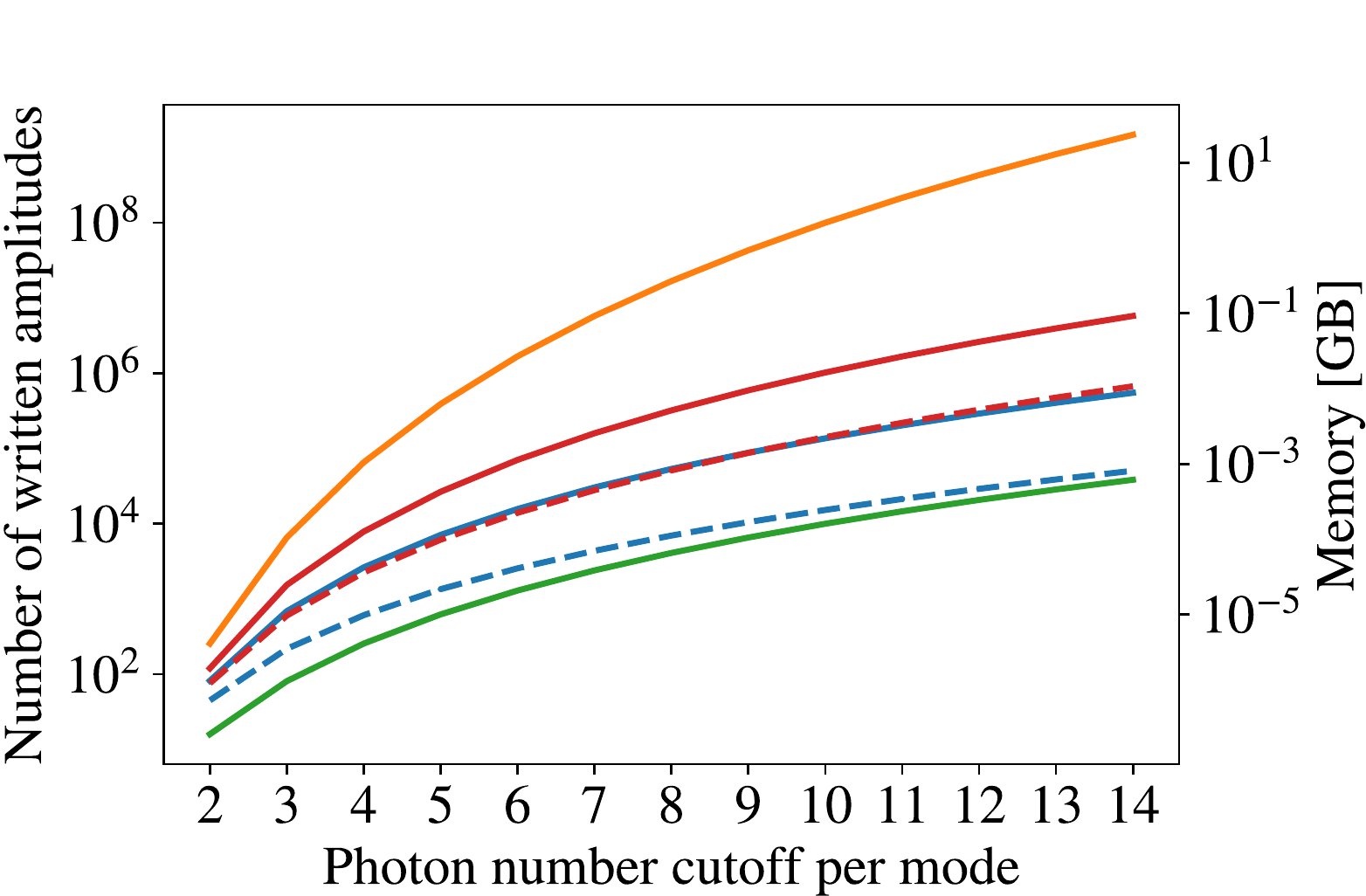}
      \caption{Memory usage}
      \label{fig:M4_space}
    \end{subfigure}
    \hspace*{\fill}%
    \begin{subfigure}{.48\linewidth}
      \centering
      \includegraphics[width=\linewidth]{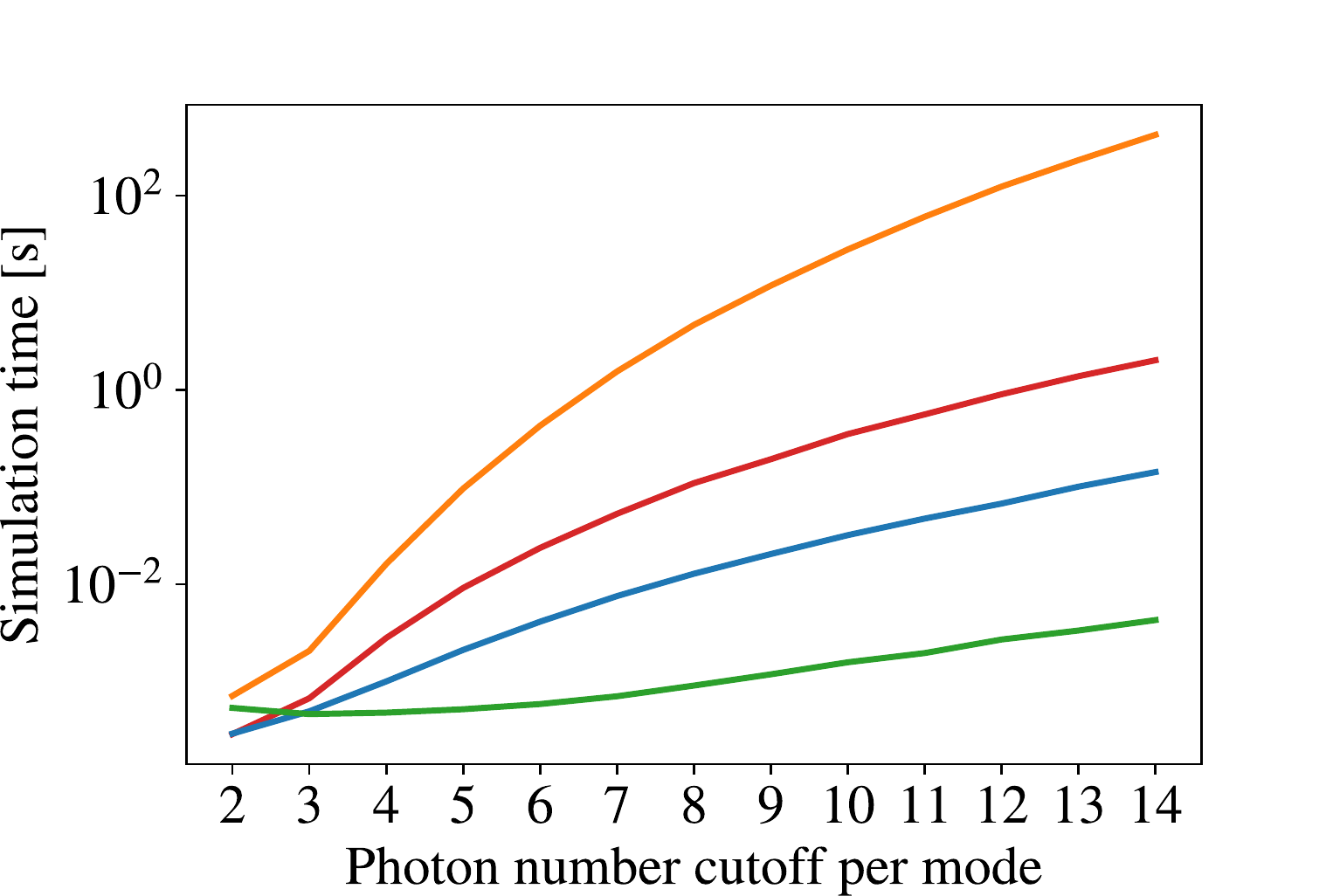}
      \caption{Simulation time}
      \label{fig:M4_time}
    \end{subfigure}
    \caption{The memory usage and simulation time of simulating a 4 mode circuit, where the photon number cutoff $C$ is chosen equal in all modes.
    The complexities of these algorithms are given by the slopes at large $C$.  Memory and time required for \cref{algo: detectAllModes pivotPlacement} scales as $C^M$, which matches the scaling of a state vector simulation (green). This is quadratic improvement over the $C^{2M}$ complexity of the naive strategy without selective pivot placement (orange). Similarly, \cref{algo: leftoverMode pivotPlacement} scales as $C^{M+1}$.
    The dashed curves show how the memory requirements of \cref{algo: detectAllModes pivotPlacement,algo: leftoverMode pivotPlacement} are lowered by the buffer strategy of \cref{sec: buffer strategy}. Note that the atypical behavior of simulation time for small cutoff is a result of time incurred during initialization.}
    \label{fig:M4_space_and_time}
    \end{figure*}

\cref{fig:M4_space_and_time} visualizes the memory usage and simulation time for a circuit with 4 modes (such as the circuits in \cref{fig:example_circuits}). We have chosen the photon number cutoff $C$ to be equal for all modes.
As both the memory usage and simulation time scale with the number of applications of \cref{eq:recrel} (i.e.~the number of pivots), the trends in \cref{fig:M4_space,fig:M4_time} are similar. 

When using state vectors, we calculate $C^M$ amplitudes to simulate a circuit, regardless of the number of PNR detectors (green line in \cref{fig:M4_space}). 
When using density matrices, this number would increase to $C^{2M}$ (orange line in \cref{fig:M4_space}) if we naively applied the strategy of \cref{sec:state_vectors}. When all modes in the circuit are measured, \cref{algo: detectAllModes pivotPlacement} reduces the memory requirements from the orange curve to the solid blue curve. This last curve corresponds with the number of written amplitudes given in \cref{sec:num written amps (detectAllModes)}. It can be lowered further to the dashed blue curve when the buffer strategy of \cref{sec: buffer strategy} is applied. Note that the memory usage at a cutoff value of 10 corresponds with the maximum of the orange curve in \cref{fig:detectAllModes_M4_bufferOverTime} .
From the slopes of these curves we verify that the complexity of \cref{algo: detectAllModes pivotPlacement} is equal to the complexity of a state vector simulation, i.e.~$C^M$, as was discussed in \cref{sec: algo1}.
When all but one mode of the circuit are detected, we can use \cref{algo: leftoverMode pivotPlacement} to improve on the naive strategy without selective pivot placement. As discussed in \cref{sec: leftover algo}, the complexity of this last algorithm is $C^{M+1}$.

When calculating both the required amplitudes (\cref{eq:recrel}) and gradients (\cref{eq:recrel_dB,eq:recrel_dA}) to optimize the circuit, we know from \cref{sec:gradients} that we can implement all three equations by taking a single walk through the Fock lattice.
As a result, the memory usage of an \textit{optimization} is a factor $1+D+D^2$ higher than the memory usage of a \textit{simulation} (where $D=M$ for state vectors and $D=2M$ for density matrices). 
When we would calculate both amplitudes and gradients for \cref{fig:M4_space_and_time} (where $M=4$), this means that the orange, red and blue curves would shift up on the log scale corresponding with a factor of $1+2M+4M^2=73$, while the factor for the green curve would be $1+M+M^2=21$. 

Note that when performing an optimization using our technique, the cost function $L$ (which could be chosen to be the fidelity to a target state for example) determines only the complexity of the first step of the chain rule, which consists in calculating $\partial L / \partial \mathcal{G}_{\bm{k}}^*$ (cf. \cref{eq:dL_dB,eq:dL_dA}). This quantity can be provided to us by an automatic differentiation framework such as TensorFlow or PyTorch. The \emph{subsequent steps} of the chain rule, which are given by the gradients that we compute in Eq.~\eqref{eq:recrel_dB} and \eqref{eq:recrel_dA} are independent of $L$ and essentially dictate the required computation time and memory resources.

\subsection{Comparison with the state-of-the-art GBS algorithm}
\label{sec:comparison_bristol}
In this section we focus our attention on the case where all modes are detected. This provides us with a useful reference point for our algorithms, since classical GBS algorithms are well studied \cite{GBS1,GBS2,quesada2019franck,bulmer2022boundary}.
We should note that there exist approximate GBS sampling algorithms such as \cite{oh2023tensor} that vastly outperform approaches where the probabilities are computed exactly in terms of simulation time and memory requirements. These are appropriate for instance in applications where the samples are needed rather than the exact probabilities. Such algorithms are not considered here.

When using state vectors, the state-of-the-art classical GBS algorithm \cite{bulmer2022boundary} obtains the probability of a single detection pattern $\bm{n}=[n_1,n_2,...,n_M]$ with a complexity that is upper bounded by $N^3 2^{N/2}$ (where $N=\sum_i n_i$) and lower bounded by $N^3 \prod_{i=1}^M \sqrt{n_i+1}$.
This algorithm is primarily used to \textit{generate samples} from a GBS circuit, i.e.~to draw a pattern $\bm{n}$ from its measurement probability distribution. A popular method for this is `chain rule sampling', where the photon number in each mode is sampled sequentially, conditioned on the photon numbers in the previous modes. This method only requires the calculation of the conditional probability distributions of the modes instead of the total joint probability distribution.

Instead of \textit{sampling} from a GBS circuit, here we obtain its \textit{joint probability distribution} by calculating the probabilities of all detection outcomes up to a certain photon number cutoff.
This is useful to study quantum algorithms based on GBS \cite{bromley2020applications,banchi2020training}.
Naively, one could apply the algorithm of Reference \cite{bulmer2022boundary} to all detection patterns up to a certain photon number cutoff. Assuming all probabilities can be obtained at the lower bound of the complexity, we get:
\begin{equation}
    \sum_{\bm{n}=\bm{0}}^{\cutoffs} N^3 \prod_{i=1}^M \sqrt{n_i+1}.
    \label{eq:bristol_naive_scaling}
\end{equation}
In \cref{app:scaling_naive_bristol} it is shown that this is higher than the complexity of our algorithm, which is $M^2 \prod_{i=1}^M C_i$. Note that to obtain $p(\bm{n})$ using \cref{algo: detectAllModes pivotPlacement}, we need to substitute $C_i$ by $n_i+1$.

Reference \cite{bulmer2022boundary} also provides a way to obtain all probabilities $p([n',n_2,...,n_M])$ (where $n' \in [0,1,...,C{-}1]$ and all other $n_i$ are fixed) at once, with the same complexity of obtaining only $p([C{-}1,n_2,...,n_M])$. Nonetheless, \cref{app:scaling_naive_bristol} shows that fixing $n_1$ to $C_1-1$ in \cref{eq:bristol_naive_scaling} still results in a complexity that is higher than $M^2 \prod_{i=1}^M C_i$.
Currently, the algorithm in Reference \cite{bulmer2022boundary} is not extended to include more than one `batched' mode, and hence our algorithm is faster at obtaining the total joint probability distribution of a GBS circuit. 
However, if an extension to multiple batched modes were to be made, it might improve on our algorithm when using state vectors. This forms an interesting open research question.

In the case of density matrix simulations, the complexity of our algorithm ($M^2 \prod_{i=1}^M C_i$) remains unaltered, while Reference \cite{bulmer2022boundary} presents a complexity of $N^3 \prod_{i=1}^M (n_i+1)$. Note that, although this last expression is quadratically higher than the \textit{lower bound} of their algorithm for state vectors, it denotes the \textit{actual} complexity to calculate a single probability $p(\bm{n})$. 
It follows that for $N^3 > M^2$ (e.g.~when $C_i>1$, $\forall \, i \in [1,2,...,M]$), our algorithm scales better, while it also produces the probabilities of all detection patterns with lower photon numbers. Consequently, two regimes can be defined for density matrix simulations. If $N^3 > M^2$, a possible extension of Reference \cite{bulmer2022boundary} to multiple batched modes would not improve on our algorithm. For $N^3 < M^2$ this question remains open for further study.

Regarding gradients, there exist alternative techniques such as computing gradients analytically or using the parameter shift rule. Computing analytical gradients, even with a given formula is usually slower than our technique because typically the analytic formula involves functions that are more complex than the steps of our recurrence relation. For example, the displacement gate entries in Fock representation are given by a combination of Laguerre polynomials, factorials and exponential functions \cite{cahill1969density} and although they can be derived analytically, the resulting derivative function is more complex than the few multiplications and additions required the recurrent formulation of the gradient of the displacement gate. Parameter shift rules \cite{schuld2019evaluating} require two forward passes per parameter, but they are not universal in the sense that there exists a parameter shift rule only for specific Gaussian unitaries (displacements, beamsplitters, squeezers etc). The complexity of the parameter shift is analogous to the complexity of our technique, however even though in this paper we have focused on Gaussian states, using recurrence relations for gradients works for any Gaussian object, including Gaussian unitaries and Gaussian channels \cite{yao2022recursive}.

\section{Conclusions}
We have presented an exact procedure to obtain the detection probabilities and conditional states of noisy linear optical quantum circuits with PNR detectors.
For a circuit with $M$ modes, we propose an algorithm for which the memory requirements and speed have a complexity of $\mathcal{O}(M^2 \prod_{i=1}^M C_i)$, where $C_i$ is the photon number cutoff of mode $i$. This constitutes a quadratic improvement over previous approaches.

The reduction in complexity applies to measured modes, even when we are after computing marginal states. Moreover, our methods can easily be adapted to obtain the gradients of the detection probabilities and conditional states with respect to a circuit parametrization. 

These methods are included in the open-source library 
\texttt{MrMustard} \cite{MrMustard_github}. They are written in pure Python using \texttt{Numpy} and are sped up using the just-in-time compiling capabilities of \texttt{Numba}.
This paves the way to making simulations and optimizations of realistic circuits with PNR detectors.
We expect our methods to accelerate the research on both GBS based algorithms and conditional state generation, with a particular emphasis on GKP state generation using ansatze such as the one in Fig.~\ref{fig:circuit_leftovermode}.

\section*{Acknowledgements}
Special thanks to Rachel S. Chadwick,  Sebasti\'an Duque Mesa, Peter Bienstman and Guy Van der Sande for the valuable discussions. The work of Robbe De Prins was performed in the context of the Flemish FWO project G006020N and the Belgian EOS project G0H1422N. It was also co-funded by the European Union in the Prometheus Horizon Europe project. His international mobility was made possible by the Scientific Research Committee (CWO) of Ghent University.
The work of Anuj Apte is supported by Yoichiro Nambu Graduate Fellowship courtesy of Department of Physics, University of Chicago.

\newpage

\bibliographystyle{plainnat}
\bibliography{main}

\begin{thebibliography}{31}
\providecommand{\natexlab}[1]{#1}
\providecommand{\url}[1]{\texttt{#1}}
\expandafter\ifx\csname urlstyle\endcsname\relax
  \providecommand{\doi}[1]{doi: #1}\else
  \providecommand{\doi}{doi: \begingroup \urlstyle{rm}\Url}\fi

\bibitem[Arrazola and Bromley(2018)]{Arrazola2018b}
Juan~Miguel Arrazola and Thomas~R. Bromley.
\newblock Using {Gaussian} boson sampling to find dense subgraphs.
\newblock \emph{Physical Review Letters}, 121\penalty0 (3), July 2018.
\newblock \doi{10.1103/physrevlett.121.030503}.

\bibitem[Arrazola et~al.(2018)Arrazola, Bromley, and Rebentrost]{Arrazola2018a}
Juan~Miguel Arrazola, Thomas~R. Bromley, and Patrick Rebentrost.
\newblock Quantum approximate optimization with {Gaussian} boson sampling.
\newblock \emph{Physical Review A}, 98\penalty0 (1), July 2018.
\newblock \doi{10.1103/physreva.98.012322}.

\bibitem[Banchi et~al.(2020{\natexlab{a}})Banchi, Fingerhuth, Babej, Ing, and
  Arrazola]{Banchi2020}
Leonardo Banchi, Mark Fingerhuth, Tomas Babej, Christopher Ing, and Juan~Miguel
  Arrazola.
\newblock Molecular docking with {Gaussian} boson sampling.
\newblock \emph{Science Advances}, 6\penalty0 (23), June 2020{\natexlab{a}}.
\newblock \doi{10.1126/sciadv.aax1950}.

\bibitem[Banchi et~al.(2020{\natexlab{b}})Banchi, Quesada, and
  Arrazola]{banchi2020training}
Leonardo Banchi, Nicol{\'a}s Quesada, and Juan~Miguel Arrazola.
\newblock Training {Gaussian} boson sampling distributions.
\newblock \emph{Physical Review A}, 102\penalty0 (1):\penalty0 012417,
  2020{\natexlab{b}}.
\newblock \doi{10.1103/PhysRevA.102.012417}.

\bibitem[Bourassa et~al.(2021)Bourassa, Alexander, Vasmer, Patil, Tzitrin,
  Matsuura, Su, Baragiola, Guha, Dauphinais, et~al.]{bourassa2021blueprint}
J.~Eli Bourassa, Rafael~N. Alexander, Michael Vasmer, Ashlesha Patil, Ilan
  Tzitrin, Takaya Matsuura, Daiqin Su, Ben~Q. Baragiola, Saikat Guha, Guillaume
  Dauphinais, et~al.
\newblock Blueprint for a scalable photonic fault-tolerant quantum computer.
\newblock \emph{Quantum}, 5:\penalty0 392, 2021.
\newblock \doi{10.22331/q-2021-02-04-392}.

\bibitem[Br{\'{a}}dler et~al.(2018)Br{\'{a}}dler, Dallaire-Demers, Rebentrost,
  Su, and Weedbrook]{Bradler2018}
Kamil Br{\'{a}}dler, Pierre-Luc Dallaire-Demers, Patrick Rebentrost, Daiqin Su,
  and Christian Weedbrook.
\newblock {Gaussian} boson sampling for perfect matchings of arbitrary graphs.
\newblock \emph{Physical Review A}, 98\penalty0 (3), September 2018.
\newblock \doi{10.1103/physreva.98.032310}.

\bibitem[Br{\'{a}}dler et~al.(2021)Br{\'{a}}dler, Friedland, Izaac, Killoran,
  and Su]{Bradler2021}
Kamil Br{\'{a}}dler, Shmuel Friedland, Josh Izaac, Nathan Killoran, and Daiqin
  Su.
\newblock Graph isomorphism and {Gaussian} boson sampling.
\newblock \emph{Special Matrices}, 9\penalty0 (1):\penalty0 166--196, January
  2021.
\newblock \doi{10.1515/spma-2020-0132}.

\bibitem[Bromley et~al.(2020)Bromley, Arrazola, Jahangiri, Izaac, Quesada,
  Gran, Schuld, Swinarton, Zabaneh, and Killoran]{bromley2020applications}
Thomas~R. Bromley, Juan~Miguel Arrazola, Soran Jahangiri, Josh Izaac,
  Nicol{\'a}s Quesada, Alain~D. Gran, Maria Schuld, Jeremy Swinarton, Zeid
  Zabaneh, and Nathan Killoran.
\newblock Applications of near-term photonic quantum computers: software and
  algorithms.
\newblock \emph{Quantum Science and Technology}, 5\penalty0 (3):\penalty0
  034010, 2020.
\newblock \doi{10.1088/2058-9565/ab8504}.

\bibitem[Bulmer et~al.(2022)Bulmer, Bell, Chadwick, Jones, Moise, Rigazzi,
  Thorbecke, Haus, Van~Vaerenbergh, Patel, et~al.]{bulmer2022boundary}
Jacob F.~F. Bulmer, Bryn~A. Bell, Rachel~S. Chadwick, Alex~E. Jones, Diana
  Moise, Alessandro Rigazzi, Jan Thorbecke, Utz-Uwe Haus, Thomas
  Van~Vaerenbergh, Raj~B. Patel, et~al.
\newblock The boundary for quantum advantage in {Gaussian} boson sampling.
\newblock \emph{Science advances}, 8\penalty0 (4):\penalty0 eabl9236, 2022.
\newblock \doi{10.1126/sciadv.abl9236}.

\bibitem[Cahill and Glauber(1969)]{cahill1969density}
Kevin~E. Cahill and Roy~J. Glauber.
\newblock Density operators and quasiprobability distributions.
\newblock \emph{Physical Review}, 177\penalty0 (5):\penalty0 1882, 1969.
\newblock \doi{10.1103/PhysRev.177.1882}.

\bibitem[Fukui et~al.(2022)Fukui, Takeda, Endo, Asavanant, Yoshikawa, van
  Loock, and Furusawa]{PhysRevLett.128.240503}
Kosuke Fukui, Shuntaro Takeda, Mamoru Endo, Warit Asavanant, Jun-ichi
  Yoshikawa, Peter van Loock, and Akira Furusawa.
\newblock Efficient backcasting search for optical quantum state synthesis.
\newblock \emph{Phys. Rev. Lett.}, 128:\penalty0 240503, June 2022.
\newblock \doi{10.1103/PhysRevLett.128.240503}.

\bibitem[Gerry and Knight(2005)]{gerry2005introductory}
Christopher~C. Gerry and Peter~L. Knight.
\newblock \emph{Introductory quantum optics}.
\newblock Cambridge university press, 2005.

\bibitem[Gottesman et~al.(2001)Gottesman, Kitaev, and Preskill]{GKP_original}
Daniel Gottesman, Alexei Kitaev, and John Preskill.
\newblock Encoding a qubit in an oscillator.
\newblock \emph{Phys. Rev. A}, 64:\penalty0 012310, June 2001.
\newblock \doi{10.1103/PhysRevA.64.012310}.

\bibitem[Hamilton et~al.(2017)Hamilton, Kruse, Sansoni, Barkhofen, Silberhorn,
  and Jex]{GBS1}
Craig~S. Hamilton, Regina Kruse, Linda Sansoni, Sonja Barkhofen, Christine
  Silberhorn, and Igor Jex.
\newblock {Gaussian} boson sampling.
\newblock \emph{Phys. Rev. Lett.}, 119:\penalty0 170501, October 2017.
\newblock \doi{10.1103/PhysRevLett.119.170501}.

\bibitem[Huh and Yung(2017)]{Huh2017}
Joonsuk Huh and Man-Hong Yung.
\newblock Vibronic boson sampling: Generalized {Gaussian} boson sampling for
  molecular vibronic spectra at finite temperature.
\newblock \emph{Scientific Reports}, 7\penalty0 (1), August 2017.
\newblock \doi{10.1038/s41598-017-07770-z}.

\bibitem[Jahangiri et~al.(2020)Jahangiri, Arrazola, Quesada, and
  Killoran]{Jahangiri2020}
Soran Jahangiri, Juan~Miguel Arrazola, Nicol{\'{a}}s Quesada, and Nathan
  Killoran.
\newblock Point processes with {Gaussian} boson sampling.
\newblock \emph{Physical Review E}, 101\penalty0 (2), February 2020.
\newblock \doi{10.1103/physreve.101.022134}.

\bibitem[Kruse et~al.(2019)Kruse, Hamilton, Sansoni, Barkhofen, Silberhorn, and
  Jex]{GBS2}
Regina Kruse, Craig~S. Hamilton, Linda Sansoni, Sonja Barkhofen, Christine
  Silberhorn, and Igor Jex.
\newblock Detailed study of {Gaussian} boson sampling.
\newblock \emph{Phys. Rev. A}, 100:\penalty0 032326, September 2019.
\newblock \doi{10.1103/PhysRevA.100.032326}.

\bibitem[Miatto and Quesada(2020)]{miatto2020fast}
Filippo~M. Miatto and Nicol{\'a}s Quesada.
\newblock Fast optimization of parametrized quantum optical circuits.
\newblock \emph{Quantum}, 4:\penalty0 366, 2020.
\newblock \doi{10.22331/q-2020-11-30-366}.

\bibitem[Oh et~al.(2023)Oh, Liu, Alexeev, Fefferman, and Jiang]{oh2023tensor}
Changhun Oh, Minzhao Liu, Yuri Alexeev, Bill Fefferman, and Liang Jiang.
\newblock Tensor network algorithm for simulating experimental {Gaussian} boson
  sampling.
\newblock \emph{arXiv preprint arXiv:2306.03709}, 2023.
\newblock \doi{10.48550/arXiv.2306.03709}.

\bibitem[Quesada(2019)]{quesada2019franck}
Nicol{\'a}s Quesada.
\newblock {Franck-Condon} factors by counting perfect matchings of graphs with
  loops.
\newblock \emph{The Journal of chemical physics}, 150\penalty0 (16):\penalty0
  164113, 2019.
\newblock \doi{10.1063/1.5086387}.

\bibitem[Quesada et~al.(2019)Quesada, Helt, Izaac, Arrazola, Shahrokhshahi,
  Myers, and Sabapathy]{PhysRevA.100.022341}
Nicol{\'a}s Quesada, Luke~G. Helt, Josh Izaac, Juan~Miguel Arrazola, Reihaneh
  Shahrokhshahi, Casey~R. Myers, and Krishna~K. Sabapathy.
\newblock Simulating realistic {non-Gaussian} state preparation.
\newblock \emph{Phys. Rev. A}, 100:\penalty0 022341, August 2019.
\newblock \doi{10.1103/PhysRevA.100.022341}.

\bibitem[Sabapathy et~al.(2019)Sabapathy, Qi, Izaac, and
  Weedbrook]{PhysRevA.100.012326}
Krishna~K. Sabapathy, Haoyu Qi, Josh Izaac, and Christian Weedbrook.
\newblock Production of photonic universal quantum gates enhanced by machine
  learning.
\newblock \emph{Phys. Rev. A}, 100:\penalty0 012326, July 2019.
\newblock \doi{10.1103/PhysRevA.100.012326}.

\bibitem[Schuld et~al.(2019)Schuld, Bergholm, Gogolin, Izaac, and
  Killoran]{schuld2019evaluating}
Maria Schuld, Ville Bergholm, Christian Gogolin, Josh Izaac, and Nathan
  Killoran.
\newblock Evaluating analytic gradients on quantum hardware.
\newblock \emph{Phys. Rev. A}, 99\penalty0 (3):\penalty0 032331, 2019.
\newblock \doi{10.1103/PhysRevA.99.032331}.

\bibitem[Schuld et~al.(2020)Schuld, Br{\'{a}}dler, Israel, Su, and
  Gupt]{Schuld2020}
Maria Schuld, Kamil Br{\'{a}}dler, Robert Israel, Daiqin Su, and Brajesh Gupt.
\newblock Measuring the similarity of graphs with a {Gaussian} boson sampler.
\newblock \emph{Physical Review A}, 101\penalty0 (3), March 2020.
\newblock \doi{10.1103/physreva.101.032314}.

\bibitem[Su et~al.(2019{\natexlab{a}})Su, Myers, and
  Sabapathy]{PhysRevA.100.052301}
Daiqin Su, Casey~R. Myers, and Krishna~K. Sabapathy.
\newblock Conversion of {Gaussian} states to {non-Gaussian} states using
  photon-number-resolving detectors.
\newblock \emph{Phys. Rev. A}, 100:\penalty0 052301, November
  2019{\natexlab{a}}.
\newblock \doi{10.1103/PhysRevA.100.052301}.

\bibitem[Su et~al.(2019{\natexlab{b}})Su, Myers, and
  Sabapathy]{su2019generation}
Daiqin Su, Casey~R. Myers, and Krishna~K. Sabapathy.
\newblock Generation of photonic {non-Gaussian} states by measuring multimode
  {Gaussian} states.
\newblock \emph{arXiv preprint arXiv:1902.02331}, 2019{\natexlab{b}}.
\newblock \doi{10.48550/arXiv.1902.02331}.

\bibitem[Takase et~al.(2021)Takase, Yoshikawa, Asavanant, Endo, and
  Furusawa]{PhysRevA.103.013710}
Kan Takase, Jun-ichi Yoshikawa, Warit Asavanant, Mamoru Endo, and Akira
  Furusawa.
\newblock Generation of optical {Schr\"odinger} cat states by generalized
  photon subtraction.
\newblock \emph{Phys. Rev. A}, 103:\penalty0 013710, January 2021.
\newblock \doi{10.1103/PhysRevA.103.013710}.

\bibitem[Takase et~al.(2022)Takase, Fukui, Kawasaki, Asavanant, Endo,
  Yoshikawa, van Loock, and Furusawa]{takase2022gaussian}
Kan Takase, Kosuke Fukui, Akito Kawasaki, Warit Asavanant, Mamoru Endo,
  Jun-ichi Yoshikawa, Peter van Loock, and Akira Furusawa.
\newblock {Gaussian} breeding for encoding a qubit in propagating light.
\newblock \emph{arXiv preprint arXiv:2212.05436}, 2022.
\newblock \doi{10.48550/arXiv.2212.05436}.

\bibitem[Technologies(2022)]{MrMustard_github}
Xanadu~Quantum Technologies.
\newblock {MrMustard}.
\newblock \url{https://github.com/XanaduAI/MrMustard}, 2022.

\bibitem[Tzitrin et~al.(2020)Tzitrin, Bourassa, Menicucci, and
  Sabapathy]{PhysRevA.101.032315}
Ilan Tzitrin, J.~Eli Bourassa, Nicolas~C. Menicucci, and Krishna~K. Sabapathy.
\newblock Progress towards practical qubit computation using approximate
  {Gottesman-Kitaev-Preskill} codes.
\newblock \emph{Phys. Rev. A}, 101:\penalty0 032315, March 2020.
\newblock \doi{10.1103/PhysRevA.101.032315}.

\bibitem[Yao et~al.(2022)Yao, Miatto, and Quesada]{yao2022recursive}
Yuan Yao, Filippo~M. Miatto, and Nicol{\'a}s Quesada.
\newblock The recursive representation of {Gaussian} quantum mechanics.
\newblock \emph{arXiv preprint arXiv:2209.06069}, 2022.
\newblock \doi{10.48550/arXiv.2209.06069}.

\end{thebibliography}

% \onecolumngrid % revtex fix
\onecolumn
% \newpage
\appendix

\section{Density matrix simulation of a 3 mode GBS circuit}
\label{sec:nestedRepr_detectAllModes_M3}

\begin{figure}[H]
\centering
  \includegraphics[width=0.7\linewidth]{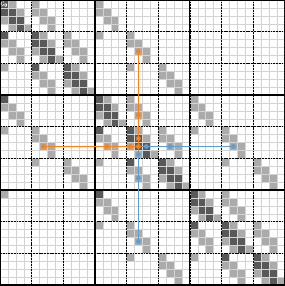}
\caption{Visual representation of \cref{algo: detectAllModes pivotPlacement} for 3 modes, with $\cutoffs=(4,3,3)$. A generalized version of \cref{fig:hypercross_M2} is used to visualize the Fock amplitudes $\mathcal{G}_{mnpqrs}$. Dark grey cells represent pivots. Light grey cells represent non-pivot amplitudes that are written. White cells represent amplitudes that are not computed. 
An animated version of this figure is included in the Supplementary Materials.}
\label{fig:nestedRepr_detectAllModes_M3}
\end{figure}

\section{Further analysis of \texorpdfstring{\cref{algo: detectAllModes pivotPlacement}}{Algorithm \ref{algo: detectAllModes pivotPlacement}}}
\label{sec:analysis algo1}
\subsection{Total number of pivots}
\label{sec:num pivots (detectAllModes)}

In this section we derive the total number of pivots that are used in \cref{algo: detectAllModes pivotPlacement}. 
First consider the diagonal pivots $\diag = [a, a, b, b, c, c, ...]$. We know there are $\prod_{i=1}^M C_i$ such amplitudes that satisfy $\bm{0} \leq \diag < \cutoffs$, but we do not have to use all of them as pivots in order to obtain all diagonal amplitudes. As is clear from \cref{fig:nestedRepr_detectAllModes_M3}, we do not use the bottom right diagonal amplitude in the diagonal $C_1 \times C_1$ blocks. In other words,
we only need $(C_1-1) \prod_{i=2}^M C_i$ diagonal pivots. In a similar way, it can be seen that we use $(C_1-1) \prod_{i=2}^M C_i$ pivots of the type $\diag + \bm{1}_{1}$, $(C_2-1) \prod_{i=3}^M C_i$ pivots of the type $\diag + \bm{1}_{3}$, $(C_3-1) \prod_{i=4}^M C_i$ pivots of the type $\diag + \bm{1}_{5}$, etc. In general, the number of pivots is equal to:
\begin{equation}
    (C_1-1) \prod_{i=2}^M C_i + \sum_{\dvar=1}^{M} \left[ (C_\dvar-1) \prod_{i=\dvar+1}^M C_i \right]~.
\end{equation}
If the photon number cutoff is equal for all modes, this simplifies to:
\begin{equation}
    (C-1)\left[C^{M-1} + \sum_{\dvar=1}^{M}C^{M-\dvar} \right] = 2C^M-C^{M-1}-1,
\end{equation}
which scales as $C^M$. 

If we use the global cutoff condition of \cref{eq:global_boundary_conditions} instead, the number of pivots is equal to:
\begin{gather*}
    \sum_{N=0}^{N_\text{max}-1}\left[\frac{(N+M-1)!}{(M-1)!N!} + \sum_{\dvar=1}^{M} \frac{(N+M-\dvar)!}{(M-\dvar)!N!} \right] \\
    = \binom{N_\text{max}+M-1}{N_\text{max}}+\binom{N_\text{max}+M}{N_\text{max}} - 1 \\
    \leq 2(N_\text{max}+M)^{\text{min}(N_\text{max},M)}-1~.
\end{gather*}

Assuming $M \ll N_\text{max}$, this upper bound scales as $(N_\text{max})^{M}$.

\subsection{Types of (off-)diagonal amplitudes}
\label{sec:off-diag types}
% Prevent duplicate writing to diagonal elements + don't write to type 0110
Let us consider the amplitudes that are read and written when applying a pivot in \cref{algo: detectAllModes pivotPlacement}. We subdivide all pivots in two types: diagonal pivots ($\diag = [a,a,b,b,c,c,...]$) and off-diagonal pivots ($\diag + \bm{1}_{\dvar}$ where $\dvar \in \{1,3,5,...,2M-1\}$).

% For simplicity we consider $M=3$ in the following example:
A diagonal pivot $diag$ reads amplitudes $\diag - \bm{1}_{i}$ and writes amplitudes $\diag + \bm{1}_{i}$, where $i \in \{1,2,3,...,2M\}$. Each amplitude of the type $\diag - \bm{1}_{i}$ can always be rewritten as  $diag' + \bm{1}_{i'}$, where $i'=i + (-1)^{i+1}$ and $diag'$ is obtained by lowering $diag_{i}$ and $diag_{i'}$ by 1. We conclude that a diagonal pivot reads and writes amplitudes of the type $\diag + \offset_1$, where $\offset_1=\bm{1}_i$ ($i \in \{1,2,3,...,2M\}$).

An off-diagonal pivot $\diag + \bm{1}_\dvar$ ($\dvar \in \{1,3,5,...,2M-1\}$) reads amplitudes $\diag + \bm{1}_\dvar - \bm{1}_i$ and writes amplitudes $\diag + \bm{1}_\dvar + \bm{1}_i$, where $i \in \{1,2,3,...,2M\}$. In a similar way it can be shown that an off-diagonal pivot only reads and writes pivots of the type $\diag + \offset$, where $\offset$ is one of the following types:
\begin{itemize}
    \item $\offset_0 = \bm{0}$
    \item $\offset_2 = 2 \cdot \bm{1}_\dvar$ ($\dvar \in \{1,3,5,...,2M-1\}$)
    \item $\offset_{1010} = \bm{1}_\dvar + \bm{1}_i$ ($\dvar \in \{1,3,5,...,2M-1\}$ and $i \in \{\dvar+2,\dvar+4,...,2M-1\}$)
    \item $\offset_{1001} = \bm{1}_\dvar + \bm{1}_i$ ($\dvar \in \{1,3,5,...,2M-1\}$ and $i \in \{\dvar+3,\dvar+5,...,2M\}$)
\end{itemize}

Note that $\offset_{0110} = \bm{1}_\dvar + \bm{1}_i$ ($\dvar \in \{1,3,5,...,2M-1\}$ and $i \in \{2,4,6,...,K-1\}$) does not occur. As is clear from line \ref{line 8} in \cref{algo: detectAllModes pivotPlacement}, off-diagonal pivots $\diag + \bm{1}_\dvar$ ($\dvar \in \{1,3,5,...,2M-1\}$) always satisfy $diag_i = 0$ for $i \in \{1,2,...,2(\dvar-1)$\}. When reading the required amplitudes for an off-diagonal pivot, indices $k_1,k_2,...,k_{2(\dvar-1)}$ therefore do not need to be lowered.

This parametrization allows all calculated amplitudes to be stored in a structured way, without storing zero values for amplitudes that do not occur in \cref{algo: detectAllModes pivotPlacement}. It can be shown that each amplitude in this structure is written exactly once. In other words, the structure is fully dense and there are no two pivots writing to the same position in the Fock lattice.
As is explained in \cref{sec: buffer strategy}, it can also be shown that every off-diagonal amplitude is included in the `read' group of a pivot exactly once, after which it can be removed from memory.

\subsection{Total number of written amplitudes}
\label{sec:num written amps (detectAllModes)}
From \cref{algo: detectAllModes pivotPlacement}, it is clear that a diagonal pivot writes at most $2M$ values, while an off-diagonal pivot $\diag + \bm{1}_\dvar$ writes at most $2(M-\dvar)$ amplitudes. The actual number of written amplitudes is determined by invoking the boundary condition $\textbf{k}<\cutoffs$.
Assuming the cutoffs in all modes to be equal to $C$, a deeper analysis shows that \cref{algo: detectAllModes pivotPlacement} writes the following number of amplitudes:
\begin{itemize}
    \item $(C-1)C^{M-1} + (C-2)C^{M-1} + (2M-2)(C-1)^2C^{M-2}$ of the type $\diag + \offset_1$
    \item $C^M$ of the type $\diag + \offset_0$ ($=\diag$)
    \item $(C-2)\sum_{\dvar=0}^{M-1}   C^{M-\dvar-1}$ of the type $\diag + \offset_2$
    \item $(C-1)^2\sum_{\dvar=0}^{M-1}(M-\dvar-1)C^{M-\dvar-2}$ of the type $\diag + \offset_{1010}$
    \item $(C-1)^2\sum_{\dvar=0}^{M-1}(M-\dvar-1)C^{M-\dvar-2}$ of the type $\diag + \offset_{1001}$
\end{itemize}

As is also shown in \cref{fig:M4_space}, the total number of amplitudes scales as $C^M$. We therefore drastically reduce the memory requirements of density matrix simulations compared to the naive strategy of calculating all $C^{2M}$ Fock amplitudes. 

\section{Scaling behaviour of \texorpdfstring{\Cref{eq:bristol_naive_scaling}}{Equation \ref{eq:bristol_naive_scaling}}}
\label{app:scaling_naive_bristol}
In this section we show that our state vector algorithm is faster than the algorithm of Reference \cite{bulmer2022boundary} at calculating the probabilities of all PNR outcomes (up to a certain cutoff) of a GBS circuit. The complexity of our algorithm is given by $M^2 \prod_{i=1}^M C_i$. If we apply the algorithm of Reference \cite{bulmer2022boundary} to all detection patterns (without using the batched strategy that was discussed in \cref{sec:comparison_bristol}), then its complexity is lower bound by \cref{eq:bristol_naive_scaling}. We assume the cutoff conditions to be constant for all modes and put $\C = C{-}1$ for ease of notation.
\cref{eq:bristol_naive_scaling} is then further lower bound by:
\begin{gather}
    \sum_{\bm{n}=\bm{0}}^{\cutoffs} N^3 \label{eq:lowerlowerbound}
    = \sum_{N=0}^{\C M} \binom{N+M-1}{N} N^3~ \\
    = \frac{\C M (\C M + 1) (\C^2 M^4 + 3 \C^2 M^3 + \C (2 \C + 3) M^2 + (3 \C - 1) M + 1) \binom{\C M + M}{\C M + 1}}{(M + 1) (M + 2) (M + 3)}~ \\
    \geq \frac{\C^4 M^6 \binom{\C M + M}{\C M + 1}}{(M + 3)^3} \geq \frac{M^6 \C^{M+3}}{(M+3)^3} \propto M^3 C^{M+3}~.
\end{gather}

If we do apply the batched strategy for one of the modes, \cref{eq:bristol_naive_scaling} can be lowered by fixing the value of $n_1$ to $\C_1 = C_1 - 1$. Consequently, the left part of \cref{eq:lowerlowerbound} becomes:
\begin{equation}
    \sum_{\bm{n}=\bm{0}}^{[\C_2,\C_3,...,\C_M]} (\C_1+N')^3 .
\end{equation}
Assuming the cutoffs to be constant in all modes, we find in a similar way that this is equal to:
\begin{gather}
    \sum_{N'=0}^{\C(M-1)} \binom{N'+M-2}{N'} (N')^3 
    \geq \frac{(M-1)^6 \C^{M+2}}{(M+2)^3} \propto M^3 C^{M+2}~,
\end{gather}
where $N' = \sum_{i=2}^M n_i$.

\end{document}